\documentclass[fleqn,usenatbib]{mnras}

\usepackage{newtxtext,newtxmath}

\usepackage[T1]{fontenc}
\usepackage{xspace}

\DeclareRobustCommand{\VAN}[3]{#2}
\let\VANthebibliography\thebibliography
\def\thebibliography{\DeclareRobustCommand{\VAN}[3]{##3}\VANthebibliography}

\newcommand{\COBEF}{{\it COBE/FIRAS}\xspace}
\newcommand{\PIXIE}{{\it PIXIE}\xspace}
\newcommand{\Planck}{{\it Planck}\xspace}

\newcommand{\LiteBIRD}{{\it LiteBIRD}\xspace}

\usepackage{graphicx}	
\usepackage{amsmath}	

\usepackage[english]{babel}
\usepackage{graphicx}
\usepackage{latexsym}
\usepackage{floatflt}
\usepackage{float}
\usepackage{units}
\usepackage{hyperref}
\usepackage{amsmath}
\usepackage{caption}
\usepackage{multicol}
\usepackage{gensymb}
\usepackage{mathdots}
\usepackage{color}
\usepackage{xcolor}
\usepackage{mathtools}
\usepackage{lipsum}
\usepackage{bm}
\usepackage{soul}

\newcommand{\id}{{\rm d}}

\title[CMB SDs and SGWBs]{Disentangling the primordial nature of stochastic gravitational wave backgrounds with CMB spectral distortions}

\author[B. Cyr et al.]{Bryce Cyr$^{1}$\thanks{Corresponding author}\thanks{E-mail: bryce.cyr@manchester.ac.uk},
Thomas Kite$^{1,2}$,
Jens Chluba$^{1}$, 
J. Colin Hill$^{3,4}$,
Donghui Jeong$^{5,6}$, \newauthor
Sandeep Kumar Acharya$^1$, 
Boris Bolliet$^{7,8}$
and Subodh P. Patil$^9$
\\
$^{1}$Jodrell Bank Centre for Astrophysics, School of Physics and Astronomy, The University of Manchester, Manchester M13 9PL, UK\\
$^{2}$Cardiology Division, Massachusetts General Hospital, Harvard Medical School, Boston, MA, USA\\
$^{3}$Department of Physics, Columbia University, New York, NY 10027, USA\\
$^{4}$Center for Computational Astrophysics, Flatiron Institute, New York, NY 10010, USA\\
$^{5}$Institute for Gravitation and the Cosmos, and the Department of Astronomy and Astrophysics,\\ \hspace{0.65mm} The Pennsylvania State University, University Park, PA 16802, USA\\
$^{6}$School of Physics, Korea Institute for Advanced Study (KIAS), 85 Hoegiro, Dongdaemun-gu, Seoul, 02455, Republic of Korea\\
$^{7}$Kavli Institute for Cosmology, University of Cambridge, Madingley Road, Cambridge CB3 0HA, UK\\
$^{8}$DAMTP, Centre for Mathematical Sciences, Wilberforce Road, Cambridge CB3 0WA, UK\\
$^{9}$Instituut-Lorentz for Theoretical Physics, Leiden University, 2333 CA Leiden, The Netherlands
}

\date{Accepted XXX. Received YYY; in original form ZZZ}

\pubyear{2023}

\begin{document}
\label{firstpage}
\pagerange{\pageref{firstpage}--\pageref{lastpage}}
\maketitle

\begin{abstract}
The recent detection of a stochastic gravitational wave background (SGWB) at nanohertz frequencies by pulsar timing arrays (PTAs) has sparked a flurry of interest. Beyond the standard interpretation that the progenitor is a network of supermassive black hole binaries, many exotic models have also been proposed, some of which can potentially offer a better fit to the data. We explore how the various connections between gravitational waves and CMB spectral distortions can be leveraged to help determine whether a SGWB was generated primordially or astrophysically. To this end, we present updated $k$-space window functions which can be used for distortion parameter estimation on enhancements to the primordial scalar power spectrum. These same enhancements can also source gravitational waves (GWs) directly at second order in perturbation theory, so-called scalar-induced GWs (SIGWs), and indirectly through the formation of primordial black holes (PBHs). We perform a mapping of scalar power spectrum constraints into limits on the GW parameter space of SIGWs for $\delta$-function features. We highlight that broader features in the scalar spectrum can explain the PTA results while simultaneously producing a spectral distortion (SD) within reach of future experiments.
We additionally update PBH constraints from $\mu$- and $y$-type spectral distortions.
Refined treatments of the distortion window functions widen existing SD constraints, and we find that a future CMB spectrometer could play a pivotal role in unraveling the origin of GWs imprinted at or below CMB anisotropy scales.
\end{abstract}




\section{Introduction}
The recent detection of a stochastic gravitational wave background (SGWB) by numerous pulsar timing array collaborations \citep{NANOGravDetection2023, EPTADetection2023, PPTADetection2023, CPTADetection2023} at nanohertz frequencies represents a major step forward in our understanding of the underlying tensor perturbations that permeate our universe. 
Unlike the initial detection of gravitational waves by LIGO \citep{LIGOScientific2016}, the progenitor of the SGWB detected in pulsar timing residuals is highly uncertain. The standard astrophysical interpretation is that such a background may be sourced by a population of super-massive black hole binaries (SMBHBs) residing at the centres of galaxies \citep{NANOGravSMBHB2023}. The current paradigm of structure formation indicates that the growth of galaxies is mediated both via accretion as well as a series of hierarchical mergers \citep{White1977}. It is well known that most galaxies host a super-massive black hole at their centre \citep{Kormendy2013,EventHorizonTelescope2019}, and in the process of merging, dynamical friction will drag the supermassive black holes towards the centre of the new galaxy. This process is known to produce SMBHBs \citep{Begelman1980}, where the gradual inspiral of the two black holes will emit gravitational waves in the nanohertz frequency regime.

The expected population of SMBHBs is uncertain, rendering the amplitude of the resultant SGWB difficult to estimate \textit{a priori}. However, the timing residual power spectral density is expected to follow a power law with spectral index $\gamma_{\rm BHB} = 13/3\approx 4.33$ \citep{Phinney2001}. In the NANOGrav analysis \citep{NANOGravDetection2023}, it was shown that a general power law fit to the observed background produces maximum likelihood values of $A_{\rm GWB} \simeq 6.4 \times 10^{-15}$, and $\gamma \simeq 3.2$. This result is roughly $3\sigma$ away from the theory prediction, introducing a mild tension between the data and this model. It should be noted that if one models dark matter fluctuations as a Fourier-domain Gaussian process (as opposed to the piecewise-constant representation considered primarily by NANOGrav) these posteriors can shift significantly. Additionally, a SGWB sourced by SMBHBs is expected to exhibit some small level of anisotropy, but so far none has been detected, leaving the SMBHB interpretation on uncertain ground \citep{NANOGrav2023Anisotropy}.

If one allows the creative liberties of a theorist, it is possible to produce this gravitational wave background in a number of extensions to the standard model, some of which have been studied by the NANOGrav collaboration directly \citep{NANOGrav2023Exotic}. Their results indicate that scalar-induced gravitational waves (SIGW) \citep{Matarrese1992, Matarrese1993, Matarrese1997, Mollerach2003, Ananda2006, Baumann2007}, first order phase transitions \citep{Kosowsky1991, Kosowsky1992a, Kosowsky1992b, Kamionkowski1993, Caprini2007, Hindmarsh2013, Hindmarsh2015}, metastable cosmic strings \citep{Buchmuller2021, Buchmuller2023}, stable cosmic superstrings \citep{Jackson2004}, and biased domain walls \citep{Kibble1976, Press1989, Vilenkin1981, Hiramatsu2010, Kawasaki2011} all provide a better fit to the SGWB than SMBHBs alone (where model comparison was performed by estimating the respective Bayes factor). This sentiment has also been echoed by subsequent analysis of the data \citep{Figueroa2023, Madge2023, Wu2023, Ellis2023}. 

Following the announcement of this discovery, a number of papers have appeared which discuss the constraints and interpretations of this SGWB in ways which modify the small-scale ($k \gtrsim 1 \, {\rm Mpc}^{-1}$) primordial power spectrum. Two well-studied examples are by sourcing the signal through SIGW \citep{Cai2023, Huang2023, Wang2023, Zhu2023,Yi2023,Yi2023b, Firouzjahi2023, You2023, Balaji2023, Yuan2023, Choudhury2023, Unal2023, Jin2023} or primordial black holes \citep{Franciolini2023, Depta2023, Inomata2023, Wang2023, Bhaumik2023, Mansoori2023, Huang2023}. Enhancements to the power spectrum on these scales can be constrained strongly by distortions to the frequency spectrum of the cosmic microwave background (CMB) \citep[e.g.,][]{Chluba2012inflaton}. Dissipation of small-scale curvature perturbations (Silk damping) mix photons with different temperatures, inducing spectral distortions even if the power spectrum remains nearly scale invariant at arbitrarily high $k$-modes. Naturally, if the scalar spectrum is enhanced on these small scales, significant distortions can be generated which allows us to derive constraints on a variety of models.

Crucially, for Gaussian scalar perturbations, \COBEF has already ruled out a primordial origin of the supermassive BHs residing in the centers of galaxies \citep{Kohri2014, Nakama2017}, which can be directly deduced from the limits first presented in \citet{Chluba2012inflaton}. The dissipation of small-scale perturbations during Big Bang Nucleosynthesis (BBN) further limits the amplitude of perturbations \citep{Jeong2014}. Various observational consequences of primordial black holes (PBHs) also present much weaker constraints at even smaller scales. 

While it will undoubtedly take some time before a precise model stands above the rest, one major question can still to be addressed: are these gravitational waves of a primordial origin? One possible way to tackle this question is by looking for $B$-mode polarization in the CMB. The \LiteBIRD collaboration will eventually be able to directly constrain very large scale tensor modes ($10^{-19} \, {\rm Hz} \lesssim f \lesssim 10^{-15} \, {\rm Hz}$) by searching for this polarization signal \citep{LiteBIRD2022} from space. Additional ground based efforts such as the Simons Observatory \citep{SimonsObservatory2018} and CMB-S4 \citep{Abazajian2019,CMB-S42020} will also probe $B$-mode polarization signals, promising stringent limits in the coming $\simeq 5-10$ years.

There is, however, a complementary way to determine if a given tensor spectrum is primordial, using CMB spectral distortions. Similar to the scalar perturbations, distortions are also generated through the direct dissipation of tensor modes present at very early times ($5\times 10^4 \lesssim z \lesssim 10^6$) \citep{Ota2014, Chluba2015, Kite2020}. In addition, models which produce the SGWB via small scale features in the power spectrum may also be probed by spectral distortions if these enhancements take place at $1 \, {\rm Mpc}^{-1} \lesssim k \lesssim 10^4 \, {\rm Mpc}^{-1}$ \citep[e.g.,][]{Chluba2012, Chluba2012inflaton, Khatri2012short2x2, Pajer2012b, Chluba2013iso, ChlubaSmall2015}. 

Within $\Lambda$CDM, a clear target of $\mu=2\times 10^{-8}$ for the $\mu$-distortion exists \citep{Chluba2012, Cabass2016, Chluba2016}, and any departure from this value would point towards new physics occurring in the pre-recombination era. This would provide significant evidence for not only a primordial source, but could also single out a small class of preferred exotic physics explanations. In this work, we further explore the synergies between GWs and CMB spectral distortions, highlighting how to properly leverage this invaluable information in a multi-messenger approach to unraveling the origin of the gravitational wave background.



In Section~\ref{sec:II}, we review the physical mechanism of spectral distortion generation from enhanced small scale power, as well as the dissipation of primordial tensor modes. We extend results in the literature by considering $y$-type distortions in addition to the $\mu$-distortions previously studied. We also provide a set of window functions for use in future computations, and showcase how broad features in the primordial power spectrum (PPS) generally strengthen constraints derived from CMB distortions. Following this, in Sections~\ref{sec:III} and \ref{sec:IV} we discuss the possible distortion signatures of scalar induced gravitational waves and primordial black hole models, which may also produce a significant SGWB. We conclude in Section~\ref{sec:V}. In what follows, we mainly take the NANOGrav results as a case study, keeping in mind that other PTA collaborations have reported qualitatively similar results.

\section{Synergies with spectral distortions} \label{sec:II}
Distortions in the frequency spectrum of the cosmic microwave background provide a sensitive probe of perturbations to the thermal history of the universe up to redshifts of roughly $z_{\rm th} \simeq 2 \times 10^6$. The inefficiency of number changing processes (Bremsstrahlung and double Compton scattering) below this redshift ensures that a blackbody spectrum cannot be recovered if departures from a thermal spectrum are present. Instead, Compton scattering drives the spectrum to a Bose-Einstein shape, characterized by a small chemical potential ($\mu$) at redshifts of $z_{\mu y} \lesssim z \lesssim z_{\rm th}$ (where $z_{\mu y} \simeq 5\times 10^4$). At still lower redshifts ($z \lesssim z_{\mu y}$), Compton scattering freezes out and any nonthermal energy injection is instead driven towards a $y$-type distortion. As the freeze out of Compton scattering is not instantaneous, energy release between $10^4 \lesssim z \lesssim 3 \times 10^5$ will also generate \textit{residual} type distortions. These distortions are not simply a superposition of $\mu$ and $y$ types, but also contain non-linear scattering residuals which can be used to probe the exact time dependence of energy injection scenarios \citep[e.g.,][]{Chluba2011therm, Khatri2012mix, Chluba2013Green} and can be parameterized using the principal components of the distortion \citep{Chluba2013PCA}.

The generation of these monopole spectral distortion signatures has been the focus of study for many years \citep{Zeldovich1969, Sunyaev1970, Illarionov1974, Danese1982, Burigana1991, Hu1993}, with modern analytic \citep{Chluba2013Green, Chluba2015GreensII} and numerical tools\footnote{\url{www.Chluba.de/CosmoTherm}} \citep{Chluba2011therm} being developed to compute constraints on both standard model and exotic physics scenarios.

One such mechanism for sourcing spectral distortions is through the dissipation of primordial perturbations in the pre-recombination era \citep{Sunyaev1970mu, Daly1991, Hu1994, Chluba2012}. When present, energy stored within these perturbations is transferred to the thermal bath through electron scattering and free-streaming effects which mix blackbodies with slightly different temperatures. Assuming that the spectrum of scalar fluctuations remains nearly scale-invariant, the dissipation of power from small scale acoustic modes ($50 \, \textrm{Mpc}^{-1} \lesssim k \lesssim 10^4 \, \textrm{Mpc}^{-1}$) induces a $\mu$-distortion with amplitude $\simeq 2 \times 10^{-8}$. This is one of the most important primordial distortion signals expected within the standard concordance model of cosmology ($\Lambda$CDM) \citep{Chluba2016}, and presents a tantalizing target in reach of next-generation spectral distortion experiments \citep{Kogut2011PIXIE, Kogut2016SPIE, Chluba2021Voyage}, although mitigating low-frequency foreground contamination is a significant challenge \citep{abitbol_pixie}. Limits on the power spectrum at these scales are complementary to the tight constraints one can derive from CMB temperature anisotropies \citep{Planck2018params} at scales of $10^{-3} \, \textrm{Mpc}^{-1} \lesssim k \lesssim 1 \, \textrm{Mpc}^{-1}$.

In addition to this standard model signal, many extensions exist which invoke an enhancement of power on sufficiently small scales. In particular, several inflationary scenarios can achieve small scale enhancement by postulating interactions between multiple fields \citep{Silk1987, Polarski:1992dq, Braglia2020}, or from adjustments to the inflaton potential \citep{Ballesteros2017, Garcia-Bellido2017}. Depending on the amplitude and wavenumber of these enhancements, it may be possible to form primordial black holes (PBHs) \citep{Carr2020, Green2020}, source gravitational waves (at second order in perturbation theory, see \citet{Domenech2021} for a review), and induce significant spectral distortions \citep{Chluba2012inflaton, Nakama2017}. The detection of these spectral distortions could provide an important multi-messenger signature of a stochastic gravitational wave background present at high redshifts, providing us with a pathway to disentangle astrophysical sources from primordial ones. 

At present, the state of the art measurement of CMB spectral distortions comes from the \COBEF satellite, which reported limits on the distortion parameters of $\mu \leq 9 \times 10^{-5}$ and $y \leq 1.5 \times 10^{-5}$ \citep{Fixsen1996} at $2\sigma$. While these bounds can be strengthened by roughly a factor of 2 \citep{tris2, Bianchini2022}, the next generation of ground based \citep[TMS][]{Jose2020TMS} and balloon borne \citep[BISOU][]{BISOU} experiments will likely further strengthen the bounds on $y/\mu$ by one order of magnitude. Satellite efforts such as \PIXIE \citep{Kogut2011PIXIE, Kogut2016SPIE} or the Voyage 2050 program \citep{Chluba2021Voyage}, however, can increase the constraining power of distortions by several orders of magnitude. With current technology, a \PIXIE-type mission is expected to reach a sensitivity of $\mu \lesssim 10^{-8}$ and $y \lesssim 2 \times 10^{-8}$, even if significant uncertainties with respect to foreground removal remain \citep{Mayuri2015, Vince2015, Mashian2016, abitbol_pixie, Rotti2021, Zelko2021}. Spectral distortions are therefore uniquely situated to provide us with a deep view behind the surface of last scattering due to the impressive increase in sensitivity possible with future experiments. 

\subsection{Spectral distortions from enhanced small-scale power}
The dissipation of small scale modes is primarily caused by the diffusion of photons which gives rise to the well known damping tail of CMB anisotropies \citep{Silk1968, Planck2018params} at $\ell \gtrsim 500$. In the context of spectral distortions, photon diffusion is equivalent to the mixing of blackbodies with slightly different temperatures which subsequently induces a distortion. This effect has been known for some time \citep{Sunyaev1970, Daly1991, Hu1994}, though more recent developments have expanded our understanding by providing complementary approaches and accurate numerical treatments \citep{Chluba2012, Pajer2012b, Chluba2011therm, Khatri2011BE,Khatri2012short2x2}. For the remainder of this subsection, we consider distortions generated by adiabatic fluctuations, following closely \citet{Chluba2012inflaton,ChlubaSmall2015}, while a treatment relevant to isocurvature modes can be found in \citet{Chluba2013iso}.

\begin{figure*}
\centering 
\includegraphics[width=\columnwidth]{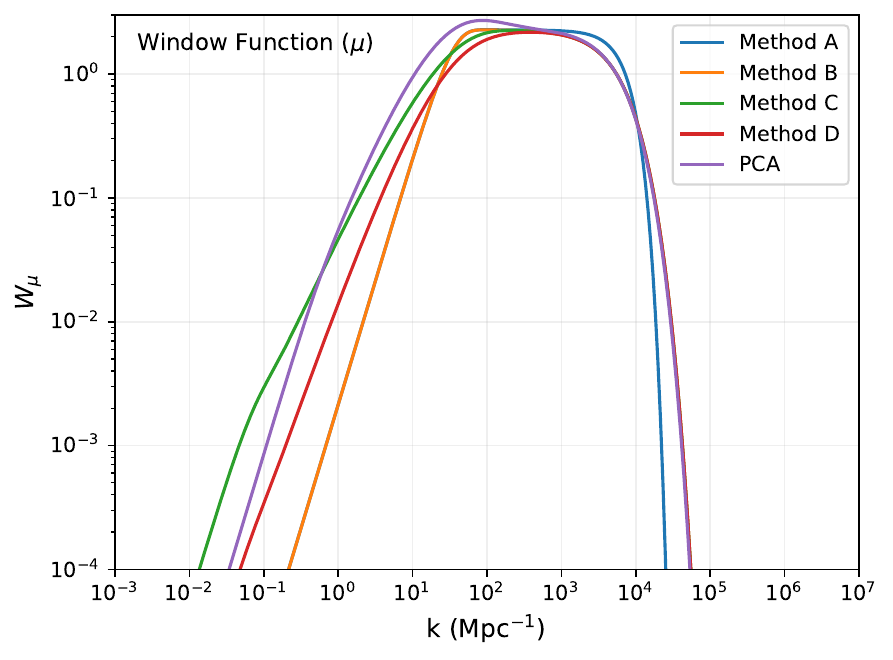}
\hspace{4mm}
\includegraphics[width=\columnwidth]{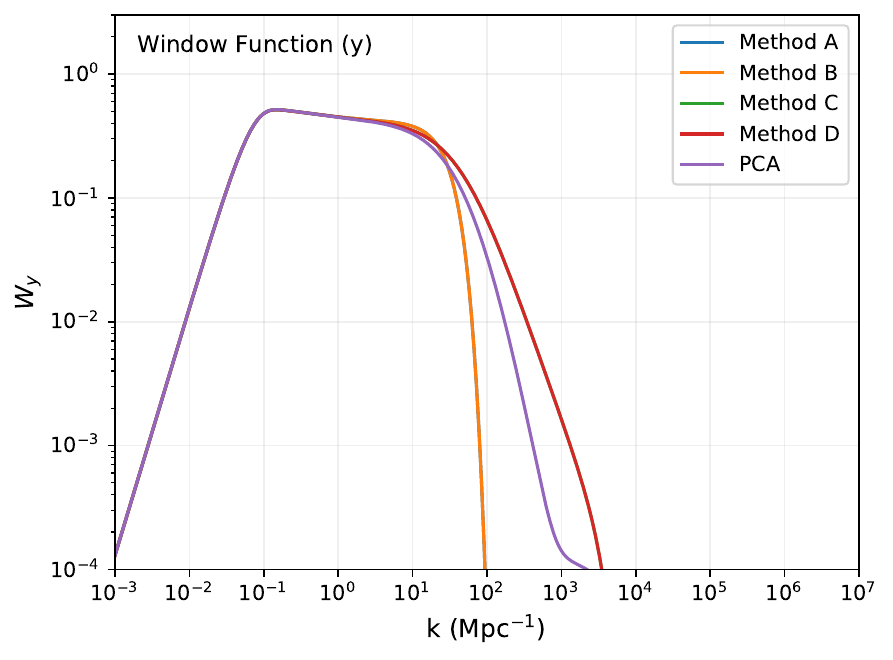}
\caption{The $k$-space window functions for $\mu$ (left) and $y$ (right) distortions. We compare the four different analytic methods discussed in the text with the numerically computed PCA projection. Method $A$ treats $z_{\rm th}$ as a hard cutoff which explains the high $k$ discrepancy for $W_{\mu}$. The other variations between methods come from different treatments of the residual era ($5 \times 10^4 \lesssim z \lesssim 3 \times 10^5$), with the PCA being the most faithful representation of the full numerical solution to the thermalization procedure, as computed in \texttt{CosmoTherm}.}
\label{fig:Window-comp}
\end{figure*}

The effective heating rate ($Q_{\rm ac}$) for this process is given by
\begin{align} \label{eq:heating-rate-1}
    \frac{\id (Q_{\rm ac}/\rho_{\gamma})}{\id z} \approx \frac{4a \dot{\tau}}{H} \int \id k \frac{k^2}{2\pi^2} &P_{\zeta}(k) \bigg[ \frac{(3 \Theta_1 - {\rm v})^2}{3} + \frac{9}{2} \Theta^2_2 \nonumber \\
    &-\frac{1}{2} \Theta_2 (\Theta_0^{\rm P} + \Theta_2^{\rm P})\bigg],
\end{align}
with $\Theta_{\rm N}$ being the Nth moment of the photon temperature or polarization transfer function (the superscript P refers to polarization), ${\rm v}$ the baryon velocity (transfer function), $P_{\zeta} = 2\pi^2 k^{-3} \mathcal{P}_{\zeta}$ the (dimensionful) primordial power spectrum, $\dot{\tau} = \sigma_{\rm T} N_{\rm e} c$ is the rate of Thomson scattering, $H$ the Hubble rate and $a$ the scale factor normalized to be unity at the present time. For primordial distortions, the tight coupling regime ensures that the quadrupole anisotropy dominates the signal, and so we neglect multipoles with $\ell \geq 2$ \citep{Chluba2012} as well as additional polarization corrections \citep{Chluba2015}, which only affect the results at the percent level. The background CMB energy density (at redshift $z$) scales as $\rho_{\gamma}\propto(1+z)^4$.

For adiabatic initial conditions and in the tight coupling regime \citep{Hu1995CMBanalytic}, it is possible to derive a more functionally useful form of this rate, namely 
\begin{align} \label{eq:heating-rate-2}
    \frac{\id (Q_{\rm ac}/\rho_{\gamma})}{\id z} \approx \frac{A^2}{Ha} \frac{32c^2}{45 \dot{\tau}(z)} \int \id k \frac{k^4}{2\pi^2} P_{\zeta}(k) {\rm e}^{-k^2/k^2_{\rm D}(z)}.
\end{align}
$A\approx 0.9$ is a dimensionless coefficient related to the neutrino loading and receives various $k$-dependent corrections in the case of isocurvature perturbations \citep{Chluba2013iso}. The damping scale $k_{\rm D}$ is determined through $\partial_{t} k_{\rm D}^{-2} \approx 8c^2/45a^2 \dot{\tau}$ \citep{Kosowsky1995, DodelsonBook} and encodes the fact that perturbations with $k > k_{\rm D}(z)$ have already dissipated their energy into the plasma at a given redshift $z$. 

Furthermore, one can find that a given $k$ mode dissipates the majority of its energy at redshift $z_{\rm diss} \approx 4.5 \times 10^5 (k/10^3 \, {\rm Mpc}^{-1})^{2/3}$. From here it is straightforward to show that $\mu$-distortions will be efficiently generated by power in modes of wavenumber $50 \, {\rm Mpc}^{-1} \lesssim k \lesssim 10^4 \, {\rm Mpc}^{-1}$, allowing us to constrain regions of the primordial power spectrum complementary to CMB anisotropy measurements.

Given the effective heating rate in Eq.~\eqref{eq:heating-rate-2}, $\mu$ and $y$-distortion estimation can proceed through a simple analytic procedure, 
\begin{align} 
\label{eq:y-estimate}
    y &= \frac{1}{4} \int_0^{\infty} \id z' \frac{\id (Q/\rho_{\gamma})}{\id z'} \mathcal{J}_y(z')
    \\
    \mu &= 1.401 \int_0^{\infty} \id z' \frac{\id (Q/\rho_{\gamma})}{\id z'} \mathcal{J}_{\mu}(z') 
    \label{eq:mu-estimate},
\end{align}
where $\mathcal{J}_{\mu/y}(z)$ are known as the distortion visibility functions, determined by the efficiency of thermalization at any given redshift \citep{Chluba2013Green,Chluba2013PCA, Chluba2016}. 

For a specified form of the visibility function, one can package the distortions in a simple way that highlights the dependence on the form of the PPS by inserting Eq.~\eqref{eq:heating-rate-2} into Eqs.~\eqref{eq:y-estimate} and \eqref{eq:mu-estimate} and computing $k$-space window functions,
\begin{align} \label{eq:final-y-estimate}
    y_{\rm ac} &\approx \int_{k_{\rm min}}^{\infty} \id k \frac{k^2}{2\pi^2} P_{\zeta}(k) W_{y}(k)\\
    \mu_{\rm ac} &\approx  \int_{k_{\rm min}}^{\infty} \id k  \frac{k^2}{2\pi^2} P_{\zeta}(k) W_{\mu} (k) \label{eq:final-mu-estimate}.
\end{align}
Here, $k_{\rm min} \approx 1 \, {\rm Mpc}^{-1}$ is a cutoff scale introduced due to the fact that modes with $k \lesssim k_{\rm min}$ are tightly constrained by CMB anisotropy, and because the efficiency of energy injection drops for larger scale modes, which dissipate in the post-recombination era through heat conduction/velocity terms rather than shear viscosity \citep{Chluba2012inflaton}. The exact form of the window function depends on our choice of distortion visibility function, of which we have five options with varying degrees of accuracy. We briefly mention each method here but refer the reader to \citet{Chluba2016} for a detailed discussion.

\textbf{Method A:} The simplest approximation one can make is to assume that the $\mu-y$ and full thermalization transitions take place instantaneously,
\begin{align}
    \mathcal{J}_{y}(z) &= \begin{cases}
    1 \hspace{4mm} {\rm for} \,\, z_{\rm rec} \leq z \leq z_{\mu y}\\
    0 \hspace{4mm} {\rm otherwise}
    \end{cases}\\
   \mathcal{J}_{\mu}(z) &= \begin{cases}
    1 \hspace{4mm} {\rm for} \,\, z_{\rm \mu y} \leq z \leq z_{\rm th}\\
    0 \hspace{4mm} {\rm otherwise}.
    \end{cases}
\end{align}
While the least accurate, this approximation can simplify the computation significantly and hence has been used in simple estimates for anisotropic distortions signals \citep[e.g.,][]{Pajer2012, Ganc2012}.

\textbf{Method B:} As a marginal improvement, one can include the fact that even for $z > z_{\rm th}$, small $\mu$-distortions can be generated by introducing the thermalization efficiency function, $\mathcal{J}_{\rm bb}(z) \approx e^{-(z/z_{\rm th})^{5/2}}$. This leads to
\begin{align}
    \mathcal{J}_{y}(z) &= \begin{cases}
    1 \hspace{4mm} {\rm for} \,\, z_{\rm rec} \leq z \leq z_{\mu y}\\
    0 \hspace{4mm} {\rm otherwise}
    \end{cases}\\
    \mathcal{J}_{\mu}(z) &= \begin{cases}
    \mathcal{J}_{\rm bb}(z) \hspace{4mm} {\rm for} \,\, z \geq z_{\rm \mu y}\\
    0 \hspace{11.3mm} {\rm otherwise}.
    \end{cases}
\end{align}
Given the wide use of this description in the literature, approximations to the $k$-space window functions have also been made \citep{Chluba2013iso, ChlubaSmall2015} for this Method
\begin{align}
    W_{y}(k) &\approx \frac{A^2}{2} {\rm exp}\left( -\left[\frac{\hat{k}}{32}\right]^2\right) \label{eq:window-B-y}\\
    W_{\mu}(k) &\approx  2.8 A^2 \left[ {\rm exp}\left(- \frac{\left(\frac{\hat{k}}{1360}\right)^2}{1+\left(\frac{\hat{k}}{260}\right)^{0.3} + \left(\frac{\hat{k}}{360}\right)} \right) - {\rm exp}\left(- \left[ \frac{\hat{k}}{32}\right]^2 \right) \right]. \label{eq:window-B-mu}
\end{align}
These can be useful for gaining analytic intuition for e.g. $\delta$-function features. Note that $\hat{k} = k/{\rm Mpc}^{-1}$.

\textbf{Method C:} We can further treat the $\mu-y$ transition redshift in a more consistent way by simply fitting the $\mu$ and $y$-distortions to their values computed numerically using a Green's function approach 
\begin{align}
    \mathcal{J}_{y}(z) &\approx \begin{cases}
    \left(1 + \left[ \frac{1+z}{6\times 10^4}\right]^{2.58} \right)^{-1} \hspace{4mm} {\rm for} \,\, z \geq z_{\rm rec} \\
    0 \hspace{29.3mm} {\rm otherwise}
    \end{cases}\\
    \mathcal{J}_{\mu}(z) &\approx  \mathcal{J}_{\rm bb} (z) \left[ 1 - {\rm exp}\left(- \left[ \frac{1+z}{5.8 \times 10^4} \right]^{1.88}  \right)\right].
\end{align}

\textbf{Method D:} The previous approach suffers from energy leakage into the residual distortions, and as a result only allows a determination of $\mu$ and $y$ to within roughly $10\%-20\%$. To enforce energy conservation, one can choose
\begin{align}
    \mathcal{J}_{y}(z) &\approx \begin{cases}
    \left(1 + \left[ \frac{1+z}{6\times 10^4}\right]^{2.58} \right)^{-1} \hspace{4mm} {\rm for} \,\, z \geq z_{\rm rec} \\
    0 \hspace{29.3mm} {\rm otherwise}
    \end{cases}\\
    \mathcal{J}_{\mu}(z) &\approx  \mathcal{J}_{\rm bb} (z)[1-\mathcal{J}_y(z)].
\end{align}

\textbf{PCA Method:} Each of the previous methods relied on peeling back layers of approximations for the treatment of the microphysics of thermalization. None of these methods, however, was able to model the complicated dynamics of the (non-$y$/non-$\mu$) residual distortion. In order to capture and utilize this information, one can resort to a principal component analysis (PCA) in which the residual part of a distortion is projected onto a basis of its eigenmodes and subsequently used to constrain energy release histories
\begin{align}
    \Delta I_i &= \Delta I_i^{\rm T} + \Delta I_i^{\mu} + \Delta I_i^{y} + \Delta I_i^{\rm Res} \\
    \Delta I_i^{\rm Res} &\approx \sum_k S_i^{(k)} \mu_k.
\end{align}

Here, the distortion signature in a narrow frequency bin $i$ can be decomposed into its contributions from the temperature shift, $\mu$-, and $y$-type spectra, plus a residual contribution. The $\mu_{k}$ are a set of distortion parameters akin to $\mu$ and $y$ which can be used to improve the accuracy of the visibility functions. Unfortunately, these improved visibility functions lack a compact analytic expression, though the interested reader can consult \citet{Chluba2013PCA}, \citet{Chluba2016} and \citet{Lucca2020} for further details. Additional parametrizations based on boosts of the $y$-distortion are discussed in \citet{Chluba2022} but will not be considered here.

For the purposes of distortions from small scale power, the window functions $W_{\mu/y}(k)$ are most relevant. Fig.~\ref{fig:Window-comp} shows a comparison of the various methods for both the $\mu$ and $y$-distortion. With the exception of Method A, the variations between procedures are most evident in the low $k$ tail of $W_{\mu}$ and the high $k$ tail of $W_y$, due to different treatments of the residual era. In addition to being the most accurate method, the PCA approach provides the most generous window function as it takes a more comprehensive distortion history into account. Below we will mainly focus on comparing Method B and the PCA, highlighting how the constraints on the small-scale power are underestimated using Method B.

\begin{figure*}
\centering 
\includegraphics[width=\columnwidth]{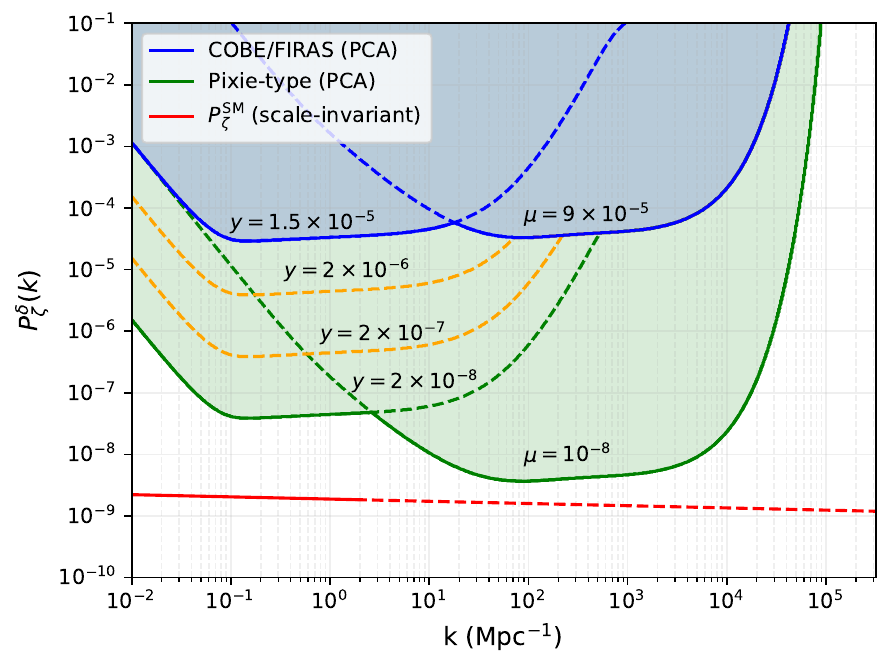}
\hspace{4mm}
\includegraphics[width=\columnwidth]{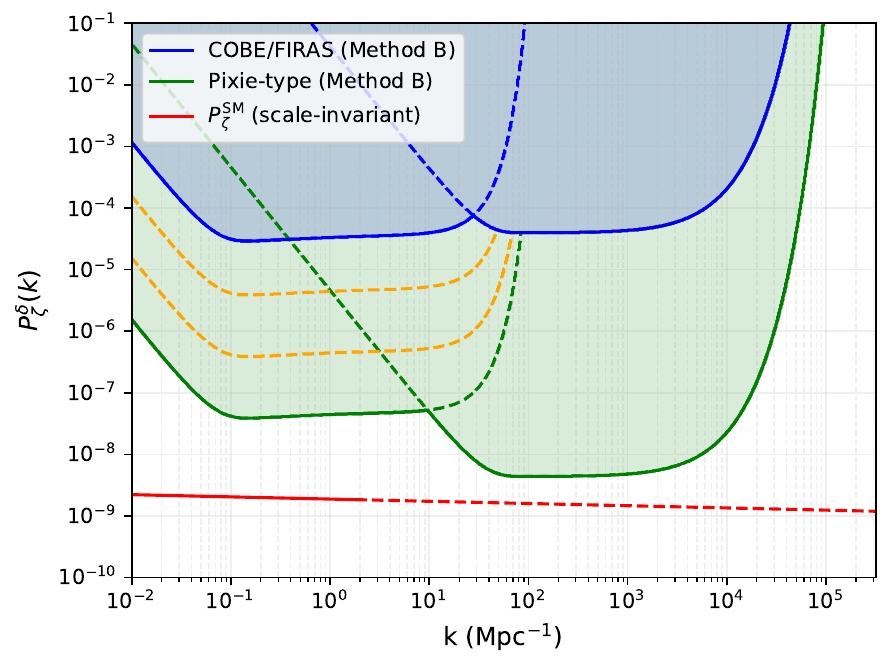}
\caption{Constraints on the amplitude ($A_{\zeta}^{\delta} = P_{\zeta}^{\delta}$) of a $\delta$-function feature in the scalar power spectrum using the full PCA window function (left), and the most commonly used analytic expressions (Method B) for $W_{\mu/y}(k)$ (right). In addition to the nominal \COBEF and \PIXIE-type distortion limits/forecasts, we also showcase a variety of intermediate limits on a next-generation primordial $y$ distortion measurement. Low redshift effects from reionization and clusters produce a $y$-distortion with amplitude $\simeq 10^{-6}$, meaning that primordial $y$ constraints will be dependent on how reliably one can marginalize out that standard model component. Constraints on broader features can be estimated by integrating the corresponding power which consequently tightens the limits, as we show in Figs.~\ref{fig:SD-constraints-Gaussian} and \ref{fig:SD-constraints-Box}.}
\label{fig:SD-constraints-PPS}
\end{figure*}

\subsection{Features in the small-scale spectrum}
At scales relevent to CMB anisotropy measurements, the spectrum of primordial scalar perturbations is well measured to be adiabatic, nearly scale-invariant, and follow Gaussian statistics \citep{Planck2018params}. The \Planck collaboration has characterized the power spectrum at scales $10^{-3} \, {\rm Mpc}^{-1} \lesssim k \lesssim 10^{-1} \, {\rm Mpc}^{-1}$ to be
\begin{align}
    \mathcal{P}_{\zeta}^{\rm SM} = A_{\zeta} \left( \frac{k}{k_{\rm p}}\right)^{n_{\rm s} - 1 + \frac{1}{2} n_{\rm run} \ln(k/k_{\rm p})},
\end{align}
with pivot $k_{\rm p} = 0.05 \, {\rm Mpc}^{-1}$, spectral index $n_{\rm s} = 0.9641$, running $n_{\rm run} = -0.0045$, and amplitude $A_{\zeta} = 2.1 \times 10^{-9}$. These correspond to the maximum likelihood values found in Tables 4 and 5 of \citet{Planck2018params} (TT,TE,EE+lowE+lensing) when adding a running of the scalar spectral index. Using Eq.~\eqref{eq:final-mu-estimate}, it is straightforward to show that if this spectrum extends to scales of $k\simeq 10^4 \, {\rm Mpc}^{-1}$, a spectral distortion with amplitude $\mu \approx 2 \times 10^{-8}$ will be generated.\footnote{The small negative contribution of $\mu=-0.3\times 10^{-8}$ from adiabatic cooling \citep{Chluba2005, Chluba2011therm, Khatri2011BE} was included in this value, implying that the acoustic damping distortion in fact totals to $\mu_{\rm ac}\simeq 2.3\times 10^{-8}$.} 
This distortion presents a target that next generation space missions such as \PIXIE\footnote{We shall use the terms \PIXIE and \PIXIE-type mission interchangeably in this work. The main targets are assessed by assuming a target of $\mu\simeq 10^{-8}$ can be achieved after foreground marginalization.} \citep{Kogut2011PIXIE, Kogut2016SPIE} or Voyage2050 \citep{Chluba2021Voyage} are capable of observing with high significance. Perhaps more exciting, however, is the possibility that these space missions could see a $\mu$-distortion with a different amplitude, as this would be a smoking gun signal of a departure from the standard $\Lambda$CDM cosmology. If one finds limits such that $\mu \lesssim 2 \times 10^{-8}$, it would signify a strong red-tilt to the small scale power spectrum, while $\mu \gtrsim 2 \times 10^{-8}$ would imply a boosted feature in the spectrum, or some other exotic energy injection mechanism such as decaying or annihilating dark matter \citep{Bolliet2020, LiuSpec2023}, cosmic strings \citep{Cyr2023, Cyr2023b}, primordial black holes \citep{Nakama2017}, or other possibilities \cite[e.g.,][]{Chluba2013fore, Chluba2013PCA}.

It is therefore possible to use spectral distortions as a tool to constrain departures from the nearly scale invariant power spectrum on small scales. Since distortion signatures are an integrated effect, constraints derived depend sensitively on the shape of the feature in the PPS. The NANOGrav collaboration has hinted that their recent detection could have been generated by some of these features at second order in perturbation theory (scalar induced gravitational waves), and that these models offer a better fit when compared against the expected astrophysical sources \citep{NANOGrav2023Exotic}. 
While the precise dynamics of GW emission from scalar perturbations poses a highly non-linear and complicated problem \citep{Pen2016}, in the perturbative regime one can convert the distortion PPS constraints into GW limits.
In following \citet{NANOGrav2023Exotic}, we thus consider spectral distortion constraints on three different shapes: $\delta$-function, Gaussian peak, and box-like features in the PPS. Additional cases are discussed in \citet{Chluba2012inflaton}.

\textbf{$\bm{\delta}$-function:} Here we assume that the feature in the power spectrum is a single $\delta$-function located at $k_*$, with amplitude $A_{\zeta}^{\delta}$
\begin{align} \label{eq:delta-source}
    \mathcal{P}_{\zeta}^{\delta} = A^{\delta}_{\zeta} \delta\big(\ln(k) - \ln(k_*)\big).
\end{align}
Depending on the position of the peak, sizeable distortions can be generated with amplitudes given by $\mu = A_{\zeta}^{\delta} W_{\mu}(k_*)$ and $y = A_{\zeta}^{\delta} W_{y}(k_*)$. In Fig.~\ref{fig:SD-constraints-PPS}, we show constraints on the amplitude of the $\delta$-function from both \COBEF (blue) as well as a \PIXIE-type satellite (green). We additionally show contours of intermediate $y$-distortion constraints. The precise {\it primordial} value of $y$ that \PIXIE will be able to constrain depends on how reliably one can subtract off the contribution coming from low redshift effects of reionization and clusters \citep[which source a $y$-distortion with amplitude $\simeq 10^{-6}$,][]{Refregier2000, Hill2015, Thiele2022}. Through upcoming direct measurements of the SZ effect of galaxy clusters and X-ray observations we expect to be able to model this contribution to the level of $1\%-10\%$, which we use here as benchmarks.

The left and right plots differ by the window function that was applied in computing the distortion. The left plot makes use of the full PCA procedure and provides stronger constraints in the range of $3 \, {\rm Mpc}^{-1} \lesssim k \lesssim 30 \, {\rm Mpc}^{-1}$ compared to the right plot. The right plot uses the Method B window function [and its analytic approximations given in Eqs.~\eqref{eq:window-B-y} and \eqref{eq:window-B-mu}] which are most commonly employed in the literature. Finally, we show the nearly scale-invariant power spectrum inferred by \Planck with the red (solid and dashed) line. While it looks like \PIXIE cannot see the extrapolated small-scale PPS signal, we remind the reader that integrating this spectrum over the window function produces $\mu \simeq 2\times 10^{-8}$, which is in reach of these next-generation experiments, assuming that the foreground challenges presented in \citet{abitbol_pixie} can be overcome.

\begin{figure*}
\centering 
\includegraphics[width=1.95\columnwidth]{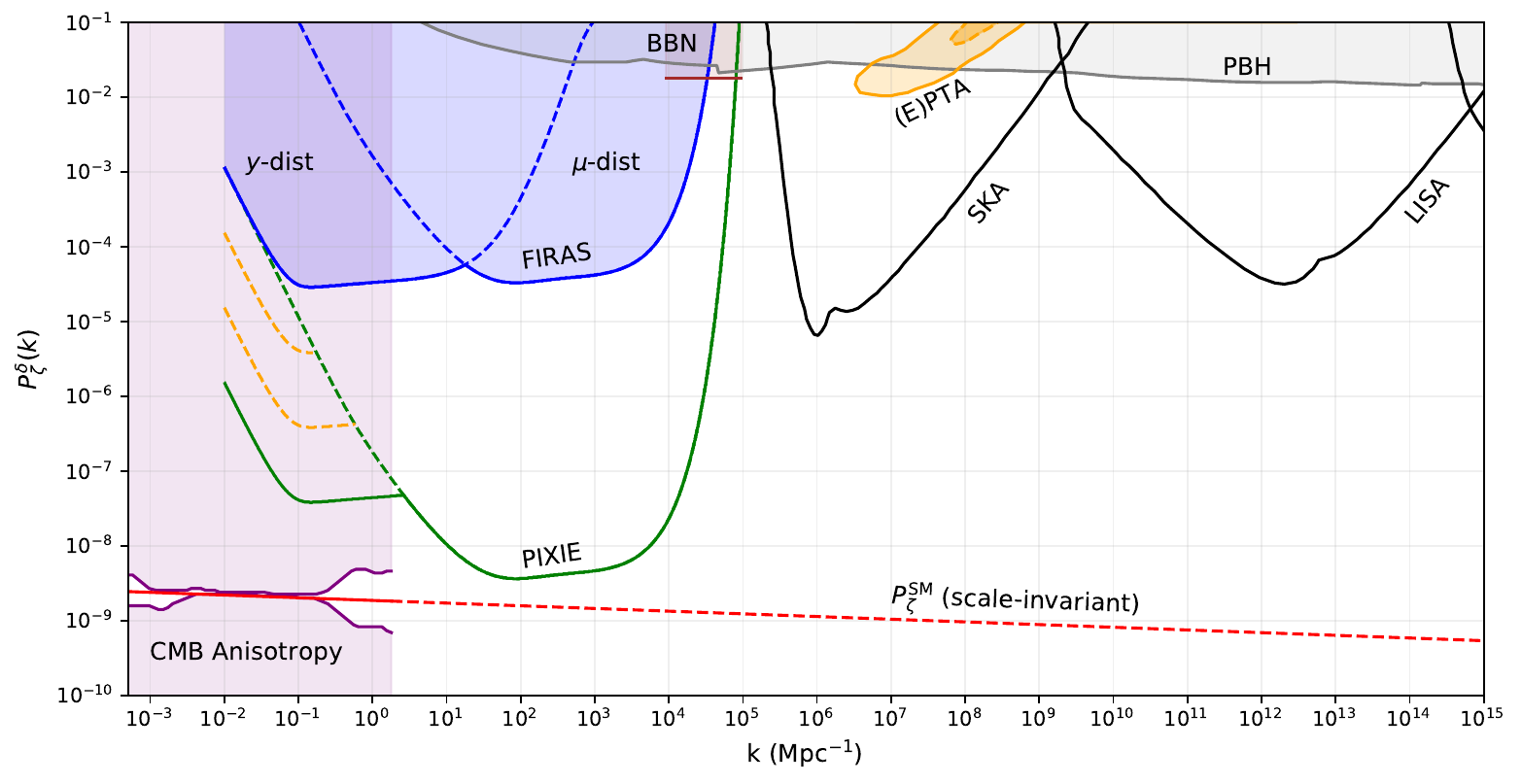}
\caption{An extended view of the primordial scalar power spectrum. The largest angular scales are tightly constrained by CMB anisotropy, with spectral distortions providing competitive bounds at scales of $1 \, {\rm Mpc}^{-1} \lesssim k \lesssim 10^4 \, {\rm Mpc}^{-1}$ \citep{Chluba2012inflaton}. At still larger scales, gravitational wave experiments can set constraints by non-detection of a stochastic background induced through SIGWs (the PTA, SKA, and LISA contours). In addition, large scalar fluctuations can seed abundant PBH production, which are constrained by various astrophysical and cosmological signatures (see \citet{Carr2010, Carr2021} and references therein for specific details). Disruptions to BBN also lead to a set of marginal constraints \citep{Jeong2014}. Current and future constraints on the spectrum are indicated by filled and unfilled contours respectively. The $1$ and $2\sigma$ detection contours reported by EPTA for a ($\delta$-function) SIGW signal are labelled in orange \citep{EPTAimplications2023}.}
\label{fig:SD-constraints-PPS-expanded}
\vspace{-3mm}
\end{figure*}

\begin{figure*}
\centering 
\includegraphics[width=\columnwidth]{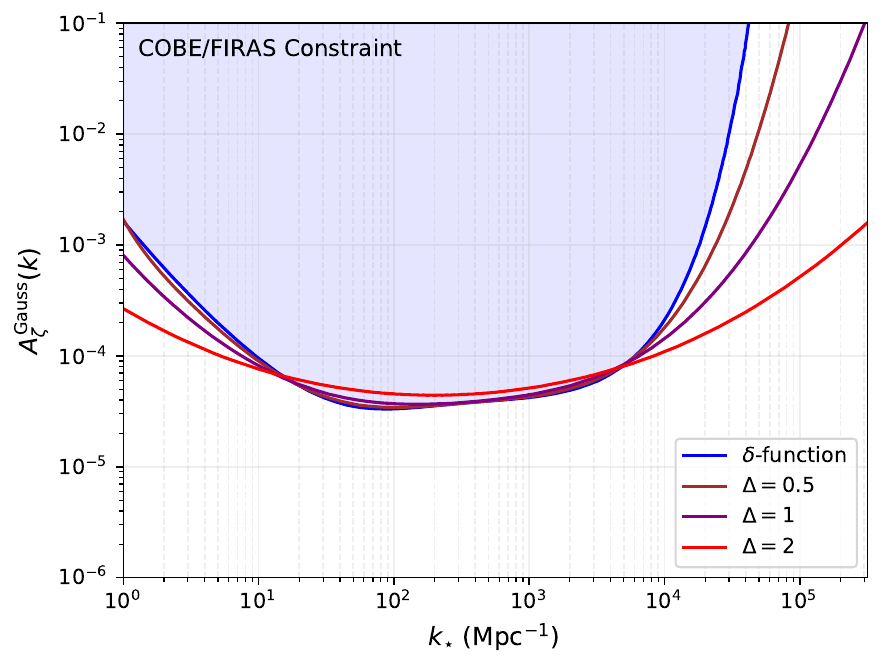}
\hspace{4mm}
\includegraphics[width=\columnwidth]{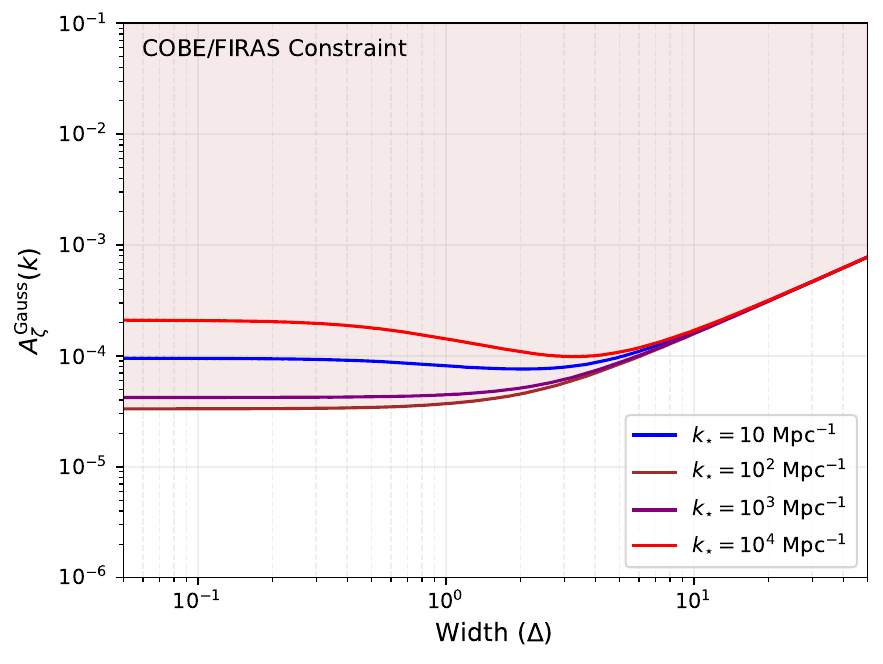}
\caption{\COBEF constraints on the amplitude of a Gaussian-type feature in the primordial power spectrum. Left: The peak position is varied for different benchmark widths. In the limit $\Delta \rightarrow 0$ the constraints approach that of the $\delta$ feature. Right: The peak positions are benchmarked and width varied.}
\label{fig:SD-constraints-Gaussian}
\end{figure*}

In Fig.~\ref{fig:SD-constraints-PPS-expanded} we show an expanded look at these constraints in the context of a broader experimental reach. On large scales precise measurements of CMB anisotropy provide stringent limits on departures from a nearly scale invariant spectrum \citep{Planck2018params}. Strong features in the power spectrum can lead to the abundant production of primordial black holes which are constrained by a variety of astrophysical and cosmological phenomena \citep{Carr2021} as indicated by the grey contour. Constraints from scalar induced gravitational waves (SIGWs) can be derived by assuming a non-detection of gravitational wave backgrounds from upcoming telescopes such as SKA, LISA, and the Einstein Telescope \citep{Gow2020, Moore2014, Bartolo2016, Maggiore2019}. Big bang nucleosynthesis (BBN) also drives constraints on the right edge of the distortion contours \citep{Jeong2014, Nakama2014, Inomata2016}. We differentiate between current and future constraints by filled and unfilled contours respectively. 

As we will discuss in the next section, the recent detection of a stochastic gravitational wave background may be sourced by a strong scalar feature. The detection contours depend sensitively on the shape of the feature, so here we choose to only show the $1$ and $2\sigma$ contours for a $\delta$-function feature as discussed in \citet{EPTAimplications2023}. In general, $\delta$ features are unphysical \citep{Cole2022}, implying that some width in $k$-space is necessary. Wide features in the power spectrum also typically boost the predicted distortion signature, which means that a \PIXIE-type experiment could potentially see a $\mu$ distortion if the detected stochastic background has a SIGW origin.
For example, the broad models illustrated in Figs.~18 and 19 of \citet{EPTAimplications2023} lead to an integrated signal of $\mu\simeq 10^{-8}-10^{-6}$ due to their enhanced power at $k\lesssim 10^5\,{\rm Mpc}^{-1}$, thus becoming visible to future CMB spectrometers, and providing a litmus test for these scenarios.

\textbf{Gaussian Peak:} For our second PPS shape, we consider departures from scale-invariance by the addition of a Gaussian peak in $\ln k$, modelled by

\begin{align}
    \mathcal{P}_{{\zeta}}^{\rm Gauss} = \frac{A^{\rm Gauss}_{\zeta}}{\sqrt{2\pi} \Delta} \exp\left[- \frac{1}{2} \left(\frac{\ln(k) - \ln(k_*)}{\Delta}\right)^2 \right].
\end{align}

This model is characterized by three parameters, and we chose to illustrate constraints on the amplitude $A^{\rm Gauss}_{\zeta}$ by benchmarking values of the width $\Delta$, and the peak position $k_*$ in Fig.~\ref{fig:SD-constraints-Gaussian}. In contrast to the $\delta$-function feature where we showcased constraints from $\mu$ and $y$, here we simply focus on $\mu$ constraints for illustrative purposes.
The left hand plot of this figure shows that the constraints become more stringent as the Gaussian widens, as even peak positions of $k_* \gtrsim 10^5 \, {\rm Mpc}^{-1}$ will have significant tails in the active regions of $W_{\mu}(k)$. As expected, the constraints approach that of a $\delta$-function in the $\Delta \rightarrow 0$ limit. 

The right hand plot shows a broader range of widths for peak positions spread throughout active areas of $W_{\mu}(k)$. For very narrow widths, the constraints become roughly constant as one approaches the $\delta$ feature, with an amplitude dependent on precisely where $k_*$ lies in the window function. For large widths, significant leakage of the signal in the high and low $k$ modes outside of the window function lead to a weakening of the constraints, coupled with an overall reduction of the effective amplitude at the peak controlled by $A^{\rm Gauss}_{\zeta}/\Delta$.

\begin{figure*}
\centering 
\includegraphics[width=\columnwidth]{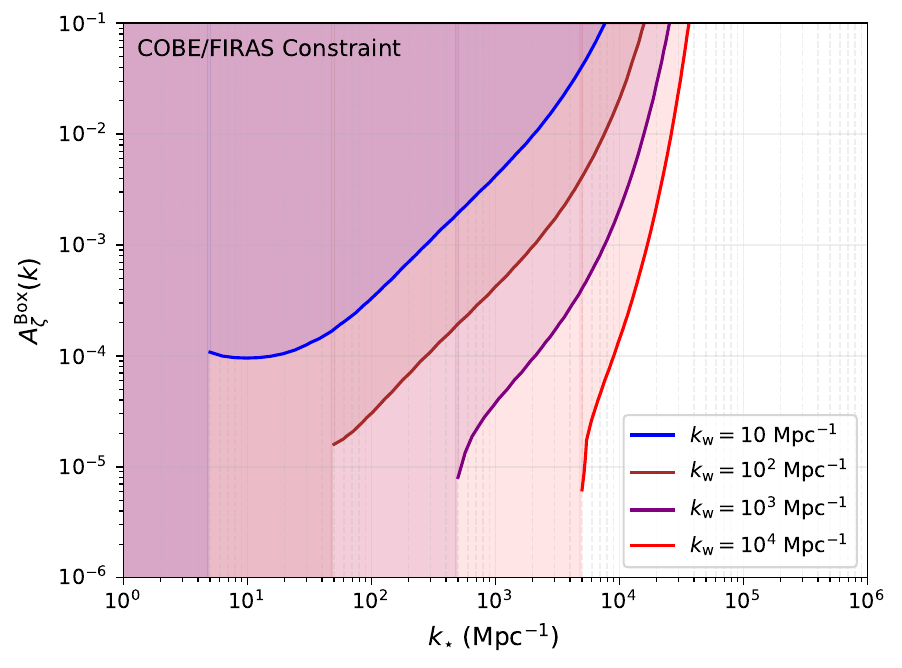}
\hspace{4mm}
\includegraphics[width=\columnwidth]{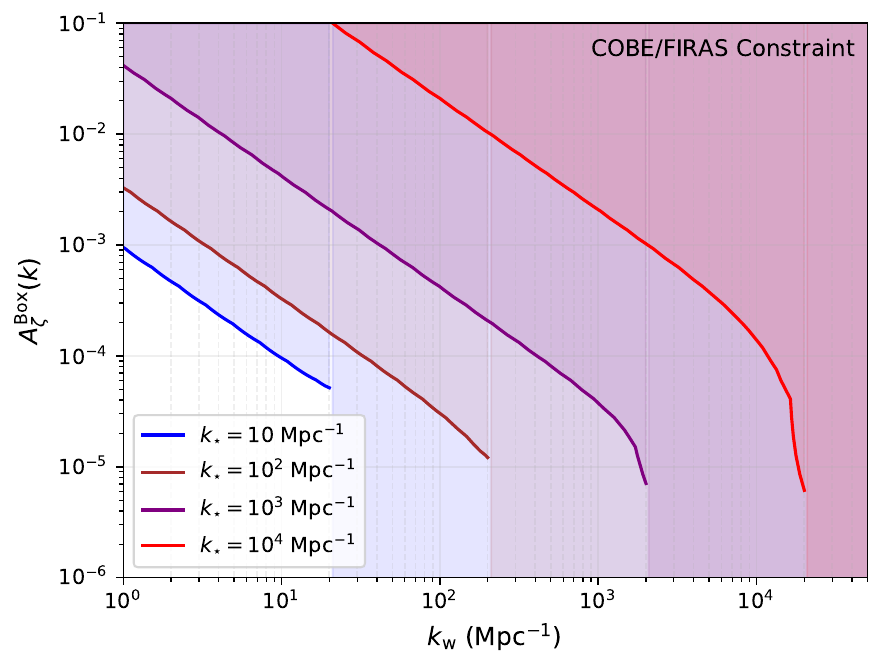}
\caption{Amplitude constraints from \COBEF for the box type feature in terms of the central position of the box $k_{*}$, and the total (linear) width, $k_{\rm w}$. Left: Peak positions are varied for different width benchmarks. The rectangular constraints on the left side of the plot are generated due to stringent constraints coming from CMB anisotropy. Right: Varying the widths for fixed peak positions. The rectangular contours on the right come from the same source as on the left. Constraints appear weaker when compared to the $\delta$-function feature for thin boxes due to an inconsistent choice of normalization, following \citet{NANOGrav2023Exotic}.}
\label{fig:SD-constraints-Box}
\end{figure*}

\textbf{Box Feature:} We additionally consider a box-like feature, also consisting of three parameters and characterized by
\begin{align}
    \mathcal{P}_{{\zeta}}^{\rm Box} = A^{\rm Box}_{\zeta}\Theta \big(\ln(k_{\rm upper}) - \ln(k)\big) \Theta \big(\ln(k) - \ln(k_{\rm lower})\big),
\end{align}
where $\Theta$ are Heaviside step functions (Not to be confused with the photon transfer functions) and $k_{\rm upper/lower}$ correspond to the upper and lower cutoffs of the box respectively. The amplitude of the $\mu$-distortion can be determined by
\begin{align}
    \mu = A^{\rm Box}_{\zeta} \int_{k_{\rm lower}}^{k_{\rm upper}} \frac{\id k}{k} W_{\mu}(k),
\end{align}
though it is more instructive to transform to parameters which characterize the central box position ($k_* = (k_{\rm lower} + k_{\rm upper})/2$), and total width ($k_{\rm w} = k_{\rm upper} - k_{\rm lower}$) when discussing constraints on the amplitude.

Similar to the Gaussian case, we show these constraints for benchmarked values of $k_{\rm w}$ and $k_*$ in the left and right plots of Fig.~\ref{fig:SD-constraints-Box} respectively. For the left hand plot, the constraints gradually strengthen for wider boxes, as expected. The block constraints on the left hand side of the plot arise due to the fact that there $k_{\rm lower} \lesssim 1 {\rm Mpc}^{-1}$ and can be strongly excluded by CMB anisotropy constraints. The right hand plot illustrates this same feature with blocked constraints on the high $k_{\rm w}$ end of the figure.

A note worth mentioning is that while the $\delta$ and Gaussian features are properly normalized to their respective amplitude parameters (in log space), the box feature is not. This leads to weaker constraints for thin boxes ($k_{\rm w} \lesssim 50 \, {\rm Mpc}^{-1}$) around $10^2 \, {\rm Mpc}^{-1} \lesssim k \lesssim 10^4 \, {\rm Mpc}^{-1}$ when compared with the $\delta$ limits. For properly normalized features, the $\delta$ feature always produces the most conservative bounds. We make this somewhat inconsistent choice of normalization to match those models considered by the NANOGrav collaboration \citep{NANOGrav2023Exotic}. To properly compare to the $\delta$ and Gaussian parameters, one should instead choose
\begin{align}
    \mathcal{P}_{\zeta}^{\rm Box, norm} = &\frac{A_{\zeta}^{\rm Box}}{\ln (k_{\rm upper}/k_{\rm lower})}\times \nonumber \\
    &\Theta \big(\ln(k_{\rm upper}) - \ln(k)\big) \Theta \big(\ln(k) - \ln(k_{\rm lower})\big).
\end{align}
In this case, taking the limit $k_{\rm w} \rightarrow 0$ will once again reproduce the most conservative $\delta$ feature constraints as expected.


\subsection{Dissipation of tensor perturbations}
Scalar perturbations are not the only type of fluctuation that can source spectral distortions at early times, both tensors (gravitational waves) and vectors offer contributions that can be significant in certain regimes. While a spectrum of vector modes sourced before the distortion window ($z \gtrsim 2 \times 10^6$) will decay quite rapidly, interesting constraints can be put on tensor modes by considering their dissipation through CMB polarization fluctuations. This process was originally studied in \citet{Ota2014,Chluba2015}, while an update regarding the complementarity of distortion based constraints and other more traditional gravitational wave observatories can be found in \citet{Kite2020} and \citet{Campeti2020}.

Tensor perturbations source spectral distortions mainly through free-streaming effects, in contrast to scalar perturbations which are damped through free-streaming as well as direct interactions with the electron-photon fluid. Indeed, it is these direct interactions (through Thomson scattering) that introduced the scalar damping scale $k_{\rm D}$ in Eq.~\eqref{eq:heating-rate-2}, efficiently converting small-scale ($k \lesssim k_{\rm D}(z)$) fluctuations into an effective heating term for the plasma. 

The lack of an efficient damping mechanism for gravitational waves introduces two qualitative difference when compared against scalars. First, the conversion from gravitational waves to an effective heating term is suppressed relative to the scalars by roughly $5$ orders of magnitude. In contrast, the second effect is that tensor perturbations can actively contribute to spectral distortions over a much wider range of scales, $1 \, {\rm Mpc}^{-1} \lesssim k \lesssim 10^6 \, {\rm Mpc}^{-1}$, with a power law decay of the window function in the UV (compared to the exponential suppression of scalars for $k \gtrsim 10^4 \, {\rm Mpc}^{-1}$).

For a given form of the primordial tensor power spectrum, the procedure to compute the associated $\mu$-distortion is very closely related to the formalism presented in Eq.~\eqref{eq:final-mu-estimate} for scalars. If the tensor spectrum is sourced at arbitrarily high redshifts (from inflation, for example), the average value of the $\mu$ distortion can be determined by
\begin{align} \label{eq:mu-tensor}
    \langle \mu_{\rm GW} \rangle =\int_0^{\infty} \id k \frac{k^2}{2\pi^2} P_{{\rm T}}(k) W_{\mu}^{\rm T}(k). 
\end{align}
In analogy with the scalar sector, $P_{\rm T}$ is the dimensionful tensor power spectrum, and $W_{\mu}^{\rm T}$ is the $k$-dependent window function for tensors. This window function is computed numerically following the details of \cite{Chluba2015}.

\begin{figure}
\includegraphics[width=\columnwidth]{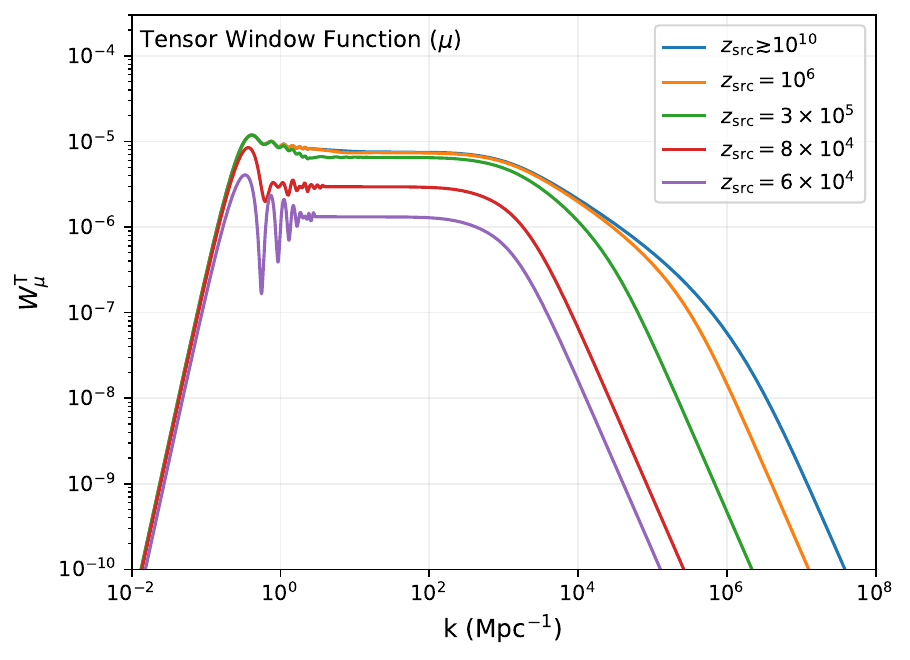}
\caption{The window function for tensor perturbations. For a long-lived gravitational wave background (blue line, $z \gtrsim 10^{10}$), the window function is maximized as tensors will dissipate over the entire $\mu$ regime. For backgrounds sourced at later times, the window function degrades as the tensors have less overall time to dissipate their energy.}
\label{fig:tensor-Wmu}
\end{figure}

There are also many scenarios in cosmology that produce a gravitational wave background significantly later than inflation, such as first order phase transitions \citep{Kosowsky1991, Kosowsky1992a, Kosowsky1992b}, cosmic strings \citep{Vilenkin1981, Vachaspati1984}, metastable domain walls \citep{Hiramatsu2010, Kawasaki2011}, and scalar induced gravitational waves \citep{Ananda2006, Baumann2007}, to name a few of interest to the NANOGrav signal. In these cases, the tensor modes will be generated over a specific redshift range on sub-Hubble scales. To account for this, \citet{Kite2020} defined a generalization of Eq.~\eqref{eq:mu-tensor}, namely
\begin{align} \label{eq:mu-tensor-general}
    \langle \mu_{\rm GW} \rangle (z = 0) =\int_0^{\infty} \id k \frac{k^2}{2\pi^2} \, \int_0^{\infty} \id z \, P_{{\rm T}}(k,z) \, \mathcal{W}_{\mu}^{\rm T}(k, z). 
\end{align}
Here, $\mathcal{W}_{\mu}^{\rm T} (k,z)$ is known as the tensor window primitive, and encodes time dependent information relevant to the damping of these more general gravitational wave spectra. In the limit where a tensor spectrum appears instantaneously, $P_{\rm T}(k,z) = P_{\rm T}(k)$, the window primitive is related to the more familiar window function via $\int_0^{\infty} \mathcal{W}_{\mu}^{\rm T} \id z = W_{\mu}^{\rm T}$. 

We show an illustration of the window functions\footnote{Window functions are available at \url{https://github.com/CMBSPEC/GW2SD.git}} in Fig.~\ref{fig:tensor-Wmu} for a primordial tensor spectrum ($z_{\rm src} \gtrsim 10^{10}$, much before the $\mu$ era), as well as for backgrounds created at later times. The amplitude of the window function is typically $5$ orders of magnitude below the scalar counterpart in the plateau region, generally leading to a weaker distortion signature except for the case of a strongly blue-tilted tensor spectrum. The amplitude of the window function decreases for tensor backgrounds created during the $\mu$ era as there is less time overall for them to dissipate their energy. For $z_{\rm src} \lesssim 5 \times 10^4$, no $\mu$ distortion is possible and so the window function closes. In principle a window function could also be constructed for $y$-distortions, though cluster marginalization makes detection of a small $y$ signal very difficult once the strong suppression relative to scalar dissipation is accounted for and so we do not consider them here.

For a more in-depth analysis of the formalism, including an explicit form for the window primitive and various mappings between $P_{\rm T}$ and the more familiar gravitational wave observable $\Omega_{\rm GW} h^2$, the reader is referred to \citet{Kite2020,Kite2021}. As indicated in Fig.~\ref{fig:SD-constraints-GWs-expanded}, spectral distortions from the direct dissipation of tensor modes are capable of probing a low frequency region of the GW parameter space not accessible by other experimental efforts. While the constraints from \COBEF are rather weak, next generation experiments will provide deep insights into tensor backgrounds over a wide range of scales. Any sufficiently strong primordial ($z_{\rm src} \gtrsim 5\times 10^4$) SGWBs will undoubtedly source spectral distortions, providing a valuable multi-messenger signal that can be useful for model discrimination.

\vspace{-3mm}
\section{Scalar induced gravitational waves} \label{sec:III}
At linear order in perturbation theory, the scalar-vector-tensor decomposition of fluctuations provides us with a powerful tool to study various phenomena. At second order, however, it is well known that scalar perturbations can source tensor modes \citep{Ananda2006, Baumann2007}. As this is a second order effect, scalar induced gravitational waves (SIGWs) suffer from significant suppression if the amplitude of the scalar power spectrum remains near its scale-invariant value at arbitrarily small scales, $\Omega_{\rm GW} \propto A_{\zeta}^2$. Therefore, theories which introduce enhancements to the small scale power will also source a spectrum of gravitational waves which can have interesting cosmological consequences.


\begin{figure}
\includegraphics[width=\columnwidth]{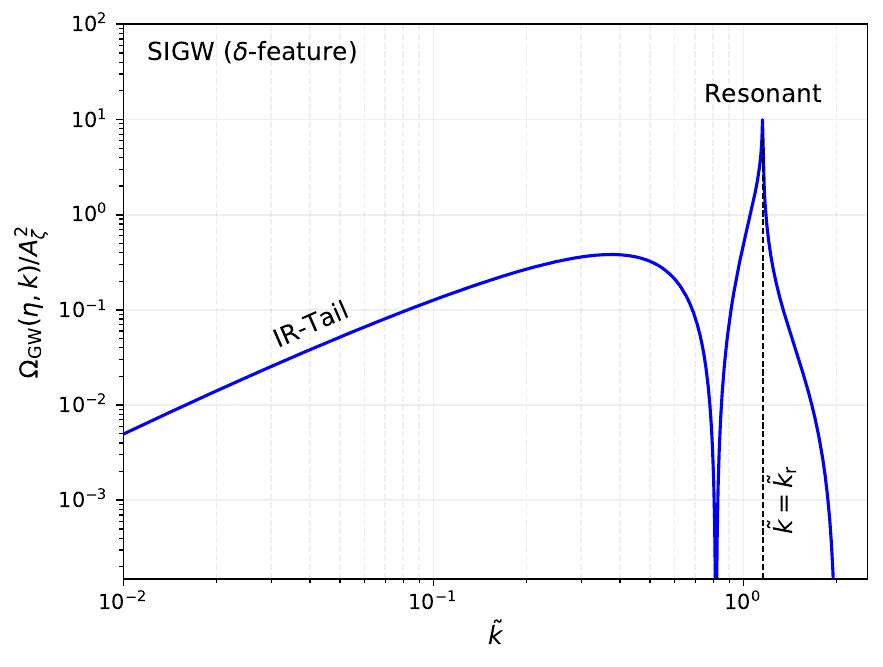}
\caption{The spectrum of induced gravitational waves from a $\delta$-function feature in the PPS. The GW resonance occurs at $\tilde{k}_{\rm r} = 2 c_{\rm s} \approx 1.15$ for tensors sourced in the radiation era.}
\label{fig:SIGW-FIRAS-detail}
\end{figure}

Perhaps the most straightforward consequence is the production of a stochastic background that could be detected by various different gravitational wave observatories. In light of the recent detection, the NANOGrav \citep{NANOGrav2023Exotic} collaboration has indicated that a fit to the data with a SIGW component is more convincing than considering SMBHBs alone. Additionally, the EPTA collaboration \citep{EPTAimplications2023} has determined $1$ and $2\sigma$ ``detection" contours for a SIGW progenitor, which we reproduced as the orange region in Fig.~\ref{fig:SD-constraints-PPS-expanded}. These contours assume a $\delta$-function feature in the scalar spectrum with an amplitude large enough that significant production of PBHs may also occur. Future constraints (assuming a non-detection) can also be forecasted for SKA, LISA, and the Einstein Telescope (see \citet{Gow2020} and references therein). The detection of this stochastic background has inspired a flurry of work on the topic \citep{Cai2023, Huang2023, Wang2023, Zhu2023,Yi2023,Yi2023b, Firouzjahi2023, You2023, Balaji2023, Yuan2023, Choudhury2023, Unal2023, Jin2023}, showcasing a broad interest in the community to further explore the parameter space of SIGWs.

Here, we would like to highlight a synergy between CMB spectral distortions and the gravitational waves induced by scalar perturbations. We choose to take a slightly different approach, where instead of converting gravitational wave detections and non-detections into constraints on the PPS (as is usually done), we ask how CMB spectral distortions and other constraints on the PPS can be transformed into limits on the gravitational wave parameter space for SIGWs.

The SIGW calculation has been refined since the seminal work of \citet{Ananda2006}, including the derivation of useful analytic forms for simple shapes in the PPS \citep{Kohri2018, Espinosa2018}. These results have been cross-checked \citep{Inomata2018} and an extensive literature now exists on the subject (see \citet{Domenech2021} for a comprehensive review). Here we summarize the results relevant to mapping the CMB spectral distortion constraints onto the gravitational wave parameter space assuming a $\delta$-function feature (i.e., the constraints presented in Figs.~\ref{fig:SD-constraints-PPS} and \ref{fig:SD-constraints-PPS-expanded}).

\begin{figure*}
\centering 
\includegraphics[width=\columnwidth]{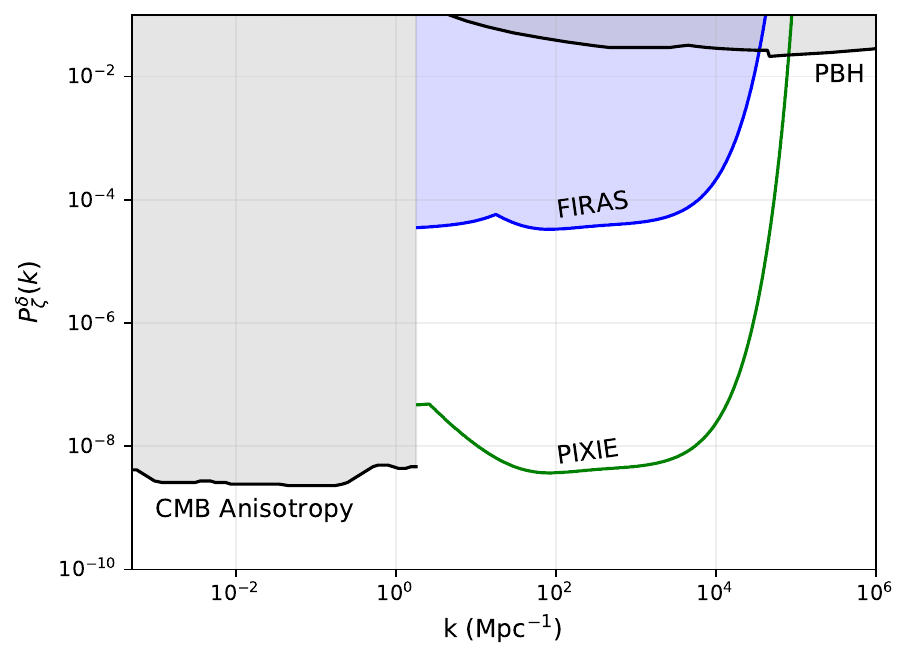}
\hspace{4mm}
\includegraphics[width=\columnwidth]{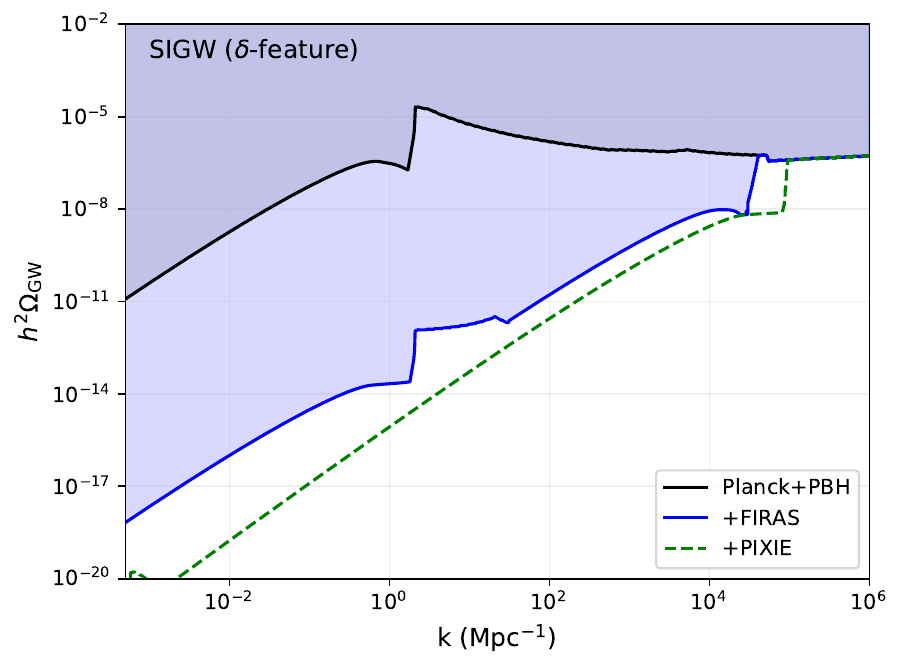}
\caption{Left: Upper limits on the primordial power spectrum for three different choices of constraints, \Planck+ PBH (black), including {\it FIRAS} (blue), and including \PIXIE (green). Right: The constrained region for the gravitational wave parameter space of SIGWs, assuming a $\delta$-function feature in the primordial power spectrum. The colour coding of the contours match the upper limits traced out on the left hand plot.}
\label{fig:SD-SIGW-comparison}
\end{figure*}

\begin{figure}
\includegraphics[width=\columnwidth]{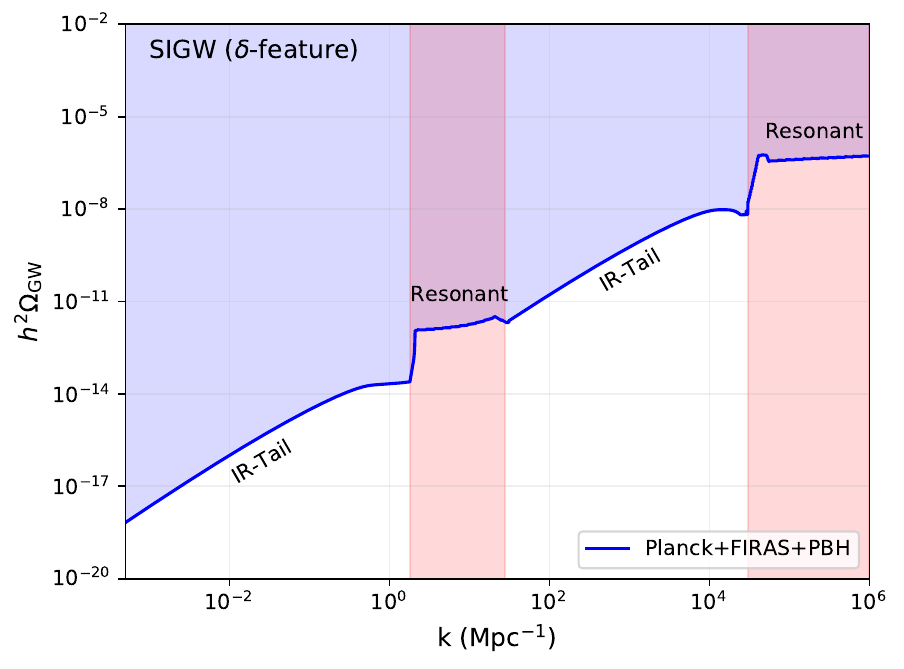}
\caption{Decomposing the FIRAS constraints into contributions from IR tails versus the resonant forests. The resonant forest constraints follow a $1 \rightarrow 1$ mapping to the limits on the PPS.}
\label{fig:SIGW-FIRAS-detail-2}
\end{figure}

We assume that the gravitational waves are induced during radiation domination. During this epoch, tensor modes are primarily sourced close to horizon re-entry for any scalar feature. Modes relevant to CMB spectral distortions generically cross the horizon in the radiation era (recall that $k_{\rm eq} \simeq 10^{-2} \, {\rm Mpc}^{-1}$). We also assume these scalar fluctuations obey Gaussian statistics. Under these simplifications, the time-averaged primordial tensor spectrum is related (at second order) to the scalar spectrum by the expression \citep{Inomata2018}
\begin{align} \label{eq:SIGW-spec}
    \overline{\mathcal{P}_{\rm h}(\eta,k)} \simeq 4 \int_0^{\infty} \id v \int_{|1-v|}^{1+v}\id u &\left( \frac{4v^2 - (1+v^2 - u^2)^2}{4 u v}\right)^2 \nonumber \\
    &\times \overline{I^2(v,u,k\eta}) \mathcal{P}_{\zeta} (kv) \mathcal{P}_{\zeta}(ku),
\end{align}
where $\eta$ is conformal time ($\id t = a \id \eta$), $v = q/k$ and $u = |\textbf{k}-\textbf{q}|/k$ relate the wavenumbers of the scalar and induced tensor modes, and $\overline{I(v,u,k\eta)}$ is a highly oscillatory kernel encoding the time-dependence of the source function (see \citet{Kohri2018, Domenech2021} for some exact expressions).

During radiation domination and in the subhorizon limit, this expression can be approximated as
\begin{align}
    \overline{I(v,u,k\eta \rightarrow \infty)} \simeq &\frac{1}{2} \left( \frac{3(u^2 + v^2 - 3)}{4u^3v^3} \frac{1}{k\eta}\right)^2 \nonumber \\
    &\times \bigg[ \bigg( - 4 u v + (u^2 + v^2 - 3) \log \left| \frac{3-(u+v)^2}{3-(u-v)^2} \right| \bigg)^2 \nonumber\\
    &+ \pi^2(u^2 + v^2 - 3)^2 \Theta(v+u - \sqrt{3}) \bigg].
\end{align}
The $\Theta$ indicate Heaviside step functions, and the gravitational wave energy density per logarithmic $k$ interval is given by
\begin{align}
    \Omega_{\rm GW}(\eta, k) = \frac{1}{24} \left( \frac{k}{a(\eta) H(\eta)} \right)^2 \overline{\mathcal{P}_{\rm h}(\eta,k)}.
\end{align}
%
For a $\delta$ feature such as the one considered in Eq.~\eqref{eq:delta-source}, the GW energy density takes a simple analytic form \citep{Kohri2018}
\begin{align} \label{eq:SIGW-delta-analytic}
    \Omega_{\rm GW}(\eta, k) &\simeq \frac{27 A^2_{\zeta}}{1024} \tilde{k}^2\left(\tilde{k}^2 - 4\right)^2 \left( \tilde{k}^2 - \frac{2}{3}\right)^2 \nonumber \\
    &\times \bigg[ 9\pi^2 \left( \tilde{k}^2 - \frac{2}{3}\right) \Theta \left( 2 \sqrt{3} - 3 \tilde{k} \right) \nonumber \\
    &+ \left( 4 + 3 \left( \tilde{k}^2 - \frac{2}{3}\right) \log \left| 1 - \frac{4}{3\tilde{k}^2}  \right| \right)^2 \bigg] \Theta\left( 2 - \tilde{k} \right),
\end{align}
where $\tilde{k} = k/k_{\rm s,*}$ (we now add the subscript $k_{\rm s}$ when discussing features related to the scalar spectrum). 

The left panel of Fig.~\ref{fig:SIGW-FIRAS-detail} illustrates the general shape of the induced GW spectrum, and consists of two interesting features: A resonant peak, occurring at $k_{\rm r} = 2c_{\rm s} k_{\rm s,*}$ ($c_{\rm s} = 1/\sqrt{3}$ in the radiation era), and an extended IR tail, which has been proposed as a fit to the NANOGrav data \citep{NANOGrav2023Exotic, You2023}. We make use of both features when performing our mapping of spectral distortion constraints from the PPS. The GW spectrum cuts off at $k \geq 2k_{\rm s,*}$ due to momentum conservation.

\begin{figure*}
\centering 
\includegraphics[width=2.05\columnwidth]{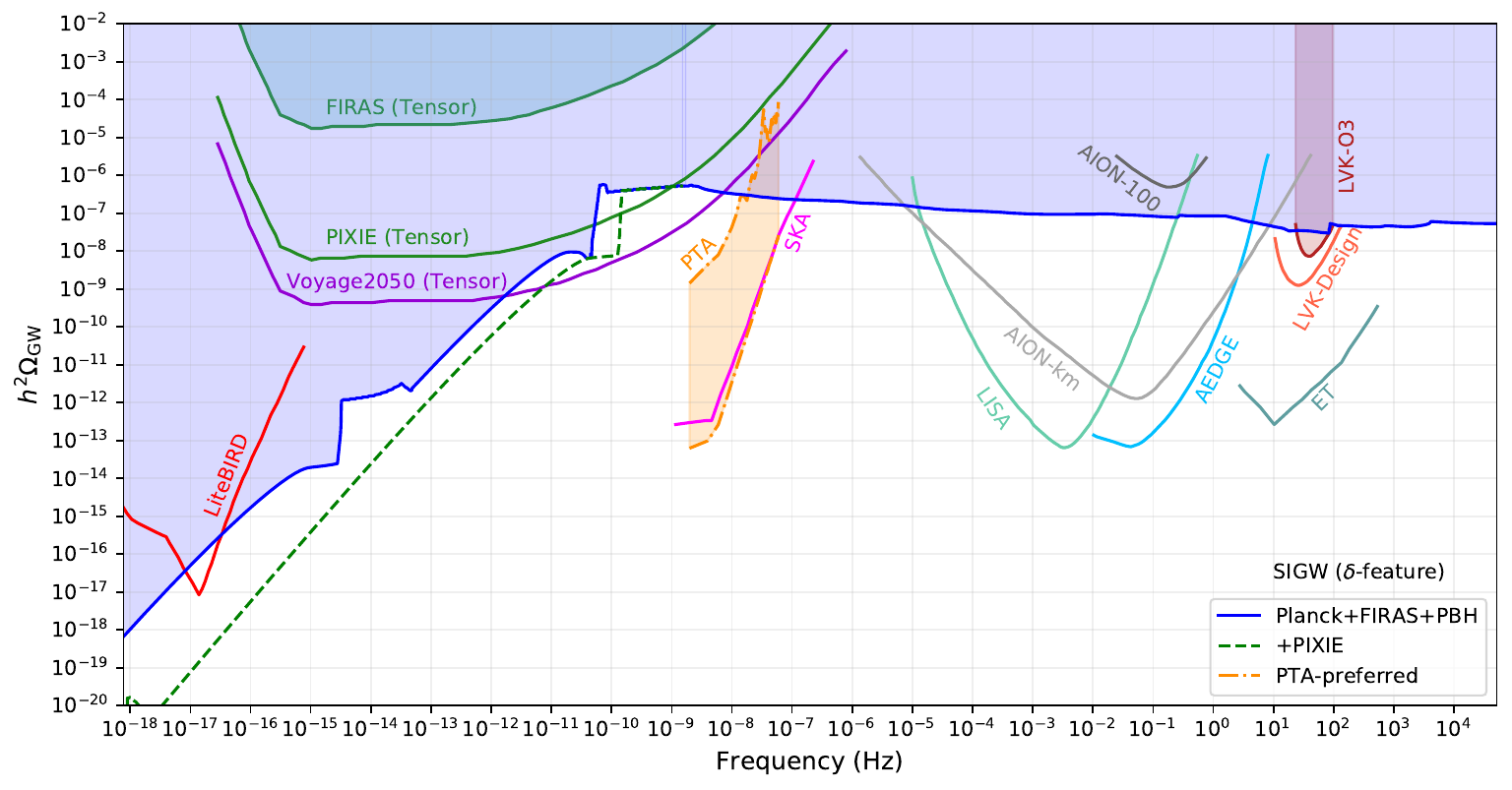}
\caption{An extended look at the gravitational wave parameter space for SIGWs from a $\delta$-function feature in the PPS. The sensitivities of various GW observatories have been compiled from \citet{Campeti2020} and \citet{Ellis2023}. Additionally, we include limits from FIRAS and future experiments on direct tensor dissipation as computed in \citet{Kite2020}. CMB spectral distortion limits on the PPS push the constraints at $f \lesssim 10^{-11} \, {\rm Hz}$ down, while PBH limits set the bound at higher frequencies. In the event that \LiteBIRD detects a primordial SGWB, a \PIXIE-type experiment would be capable of determining if they were produced by SIGWs from a $\delta$ feature. The shaded orange region represents the $2\sigma$ detection contours reported by NANOGrav \citep{NANOGravDetection2023}.}
\label{fig:SD-constraints-GWs-expanded}
\end{figure*}

After the production of these GWs, their energy density is simply redshifted to the present day for comparison with observations,
\begin{align}
    \Omega_{\rm GW}(\eta_0,k) h^2 = &1.62 \times 10^{-5} \left( \frac{\Omega_{\rm r,0} h^2}{4.18 \times 10^{-5}} \right) \nonumber \\
    &\times \left( \frac{g_*(T_{\rm c})}{106.75}\right) \left( \frac{g_{*s}(T_{\rm c})}{106.75}\right)^{-4/3} \Omega_{\rm GW}(\eta_{\rm c},k),
\end{align}
where $\eta_{\rm c}$ and $T_{\rm c}$ are the conformal time and temperature of horizon crossing for a $\delta$ feature located at $k_{\rm s}=k_{\rm s,*}$, and $\Omega_{\rm r,0}$ is the radiation density today. The usual effective spin/entropic degrees of freedom are encoded by $g_{*}$ and $g_{*s}$.

To translate the PPS constraints into limits on the SIGWs we proceed as follows: For a given value of $k_{\rm s} (=k_{\rm s,*})$, we read off the corresponding upper limit on $P_{\zeta}^{\delta}(k_{\rm s,*})$ coming from \Planck (large scales) or PBHs, \COBEF and \PIXIE (small scales) as seen in the left panel of Fig.~\ref{fig:SD-SIGW-comparison}. This amplitude is then inserted into Eq.~\eqref{eq:SIGW-delta-analytic} to infer the maximum allowed $\Omega_{\rm GW}$ at a given $k$-mode. The right panel of Fig.~\ref{fig:SD-SIGW-comparison} illustrates the evolution of these constraints when including more stringent measurements on the small scale PPS from \COBEF and a future spectral distortion mission with \PIXIE-like sensitivity.

For a \PIXIE-type experiment, the dominant constraint over modes of $10^{-2} \, {\rm Mpc}^{-1}\lesssim k \lesssim 10^4 \, {\rm Mpc}^{-1}$ comes from the IR tail of a $\delta$ feature at the PBH bound on the right hand side of the (scalar) \PIXIE contour. Therefore, it is clear that the most efficient way to strengthen constraints on the GW parameter space is by improving limits on the PPS over a wider range of scales. 

The current bounds from FIRAS can also be seen in Fig.~\ref{fig:SIGW-FIRAS-detail-2}. For small and intermediate $k$-modes, the smooth IR tails of two separate scalar $\delta$ features (located at $k_{\rm s,*} \simeq 2 \, {\rm Mpc}^{-1}$ and $k_{\rm s,*} \simeq 4\times 10^4 \, {\rm Mpc}^{-1}$) set the leading constraints. The regions marked ``resonant" are instead set by the resonance peaks of a forest of $\delta$ features and follow a $1 \rightarrow 1$ mapping onto the PPS constraints.

One subtlety regarding our mapping is related to the regularization of the resonant peak exhibited in Fig.~\ref{fig:SIGW-FIRAS-detail}. The height of this resonance is in principle unbounded, though the total integrated energy density remains finite. We choose to regulate this amplitude by demanding that the fractional energy neglected in our scheme be $\epsilon \lesssim 0.3\%$. The height of the resonance peaks, and thus our constraint contours, are dependent on the choice of $\epsilon$, so a couple of comments are in order. First, in a realistic setup, the minimum width of this resonance would be bound by finite-time effects in analogy to the finite widths of atomic spectral lines, implying that the amplitude itself is not unbounded. Secondly, gravitational wave detectors are sensitive to energy deposition in $k$-modes with a finite bin width. Our regularization condition on $\epsilon$ implies that we flatten the peak over a width of $\delta k/k_{\rm r} \simeq 10^{-4}$. Provided that this width is smaller than the binning done by any given experiment, our constraint curves will not be altered. We leave a more complete analysis of this regularization to future work.

We show an extended look at the mapping of PPS constraints on the GW parameter space by the blue shaded region in Fig.~\ref{fig:SD-constraints-GWs-expanded}. As expected, the limits we set here do not preclude a SIGW origin to the NANOGrav signal, whose preferred region of parameter space is indicated by the orange contour. Instead, we highlight that stringent constraints can be set on low frequency gravitational waves originating from $\delta$ features in the primordial power spectrum due to limits on CMB spectral distortions.

\LiteBIRD \citep{LiteBIRD2022}, CMB-S4 \citep{Abazajian2019, CMB-S42020}, and the Simons Observatory \citep{SimonsObservatory2018} are hunting for large-scale gravitational waves by searching for $B$-mode polarization in the CMB. If such a signal is detected, disentangling its possible origin will be a top priority.
For instance, if the \LiteBIRD satellite detects a signature of primordial gravitational waves, a \PIXIE-like spectral distortion experiment will possess the capability to determine whether the source of such gravitational waves is of a SIGW ($\delta$ feature) origin. This can be deduced from Fig.~\ref{fig:SD-constraints-GWs-expanded}, noting that the green dashed contour fully covers the \LiteBIRD sensitivity curve.
In addition, we can immediately conclude that \COBEF already rules out a significant SIGW contribution over most of the frequencies probed by \LiteBIRD.
Moving forward, multi-messenger signatures such as this will be crucial in determining the physical origin of various primordial processes.

The constraint contours derived in this section are robust for the case of a $\delta$ feature in the PPS. They can, however, change quite dramatically for features with more general shapes. We have chosen to show the $\delta$ constraints first as they offer an intuitive and illustrative mapping between constraints on the scalar spectrum, and those on the tensors. The obtained limits can, however, be viewed as conservative.

The mapping procedure becomes more complicated with general shapes, because both the spectral distortion limits on the PPS, and the gravitational waveform responses transform non-trivially. In general, wider shapes will induce a larger $\mu$ distortion, which in turn will lead to even more stringent constraints on the amplitude of the induced gravitational waves. We plan to explore this further and expand our results in a future work to more generic log-normal, box, and broken power-law shapes.

\section{Primordial black holes} \label{sec:IV}
The presence of enhanced density perturbations also gives rise to the possibility of PBH formation. Indeed, there have already been numerous claims of a PBH interpretation (or by-product) to the NANOGrav signal \citep{Franciolini2023, Depta2023, Inomata2023, Wang2023, Bhaumik2023, Mansoori2023, Huang2023}, as well as some work scrutinizing the idea \citep{Gouttenoire2023}. 

A population of primordial black holes can induce CMB spectral distortions in up to three unique ways. On the low mass end, PBHs in the range $10^{-20} \, M_{\odot} \lesssim M_{\rm PBH} \lesssim 10^{-17} \, M_{\odot} $ will undergo their final moments of evaporation during the distortion era, directly injecting significant amounts of energy and entropy \citep{Acharya2020, Chluba2020large}. 
Spinning PBHs undergo slightly different dynamics, but can be constrained through similar effects \citep{Pani2013}.
Over a wide range of intermediate mass scales the accretion of background material onto the PBHs can generate distortions through the creation of primordial `jet'-like features \citep{Ricotti2007}. While these distortions are too faint for \COBEF to see \citep{Horowitz2016,Ali-Haimoud2016}, there is still hope for next generation instruments, with the challenge of disentangling the $y$-type distortion from the SZ cluster contributions. Finally, the production of primordial black holes requires large features in $\zeta$, the scalar perturbation. As we have discussed above, if these features exist in a certain range of $k$-modes, they can generate sizeable spectral distortions. We review this calculation \citep[closely following][]{Nakama2017} and present updated constraints using the more precise PCA scalar window functions as shown in Fig.~\ref{fig:Window-comp}.

For simplicity, let us once again posit that the power spectrum possesses a $\delta$-function feature at some scale $k_*$, namely $\mathcal{P}_{\zeta}^{\delta} = A_{\zeta}^{\delta} \delta[\ln(k) - \ln(k_*)]$. Assuming Gaussian statistics \citep[for a non-Gaussian extension see][]{Nakama2017} the probability density function for the curvature perturbation in a given patch of the sky is given by 
\citep{Nakama2017, Carr2019}
\begin{align}
    P(\zeta) = \frac{1}{\sqrt{2\pi} \sigma} \exp\left(- \frac{\zeta^2}{2\sigma^2} \right),
\end{align}
where $\sigma$ is the dispersion of the perturbations smoothed over the horizon. For $\delta$-function features, the amplitude and dispersion are related by $A_{\zeta}^{\delta} = \sigma^2$.

Numerical simulations indicate that gravitational collapse of a Hubble patch occurs when the local curvature perturbation exceeds some critical value, $\zeta \gtrsim \zeta_{\rm c}$. The precise value of $\zeta_{\rm c}$ is still a topic of some debate, but likely lies within the $\mathcal{O}(0.1 - 1)$ range. For our purposes, we take $\zeta_{\rm c} = 0.67$ as found in \citet{Harada2017} to provide a more direct comparison with the results of \citet{Nakama2017} who used this same value.

\begin{figure}
\includegraphics[width=\columnwidth]{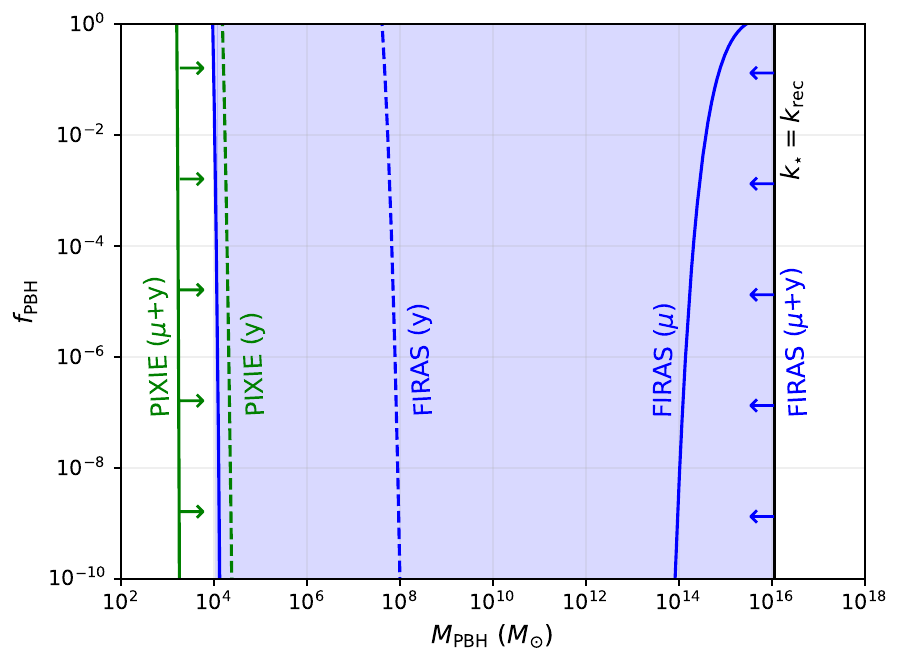}
\caption{Updated spectral distortion constraints on the $f_{\rm pbh}$-$M_{\rm pbh}$ phase space. The region already excluded by \COBEF is shown in blue, and future \PIXIE-type forecasts are shown in green. Dashed and solid lines show the $y$ and $\mu$ contours respectively. Note that the \COBEF $y$ contour extends all the way to the black line on the right, which corresponds to scales entering horizon post-recombination. Both \PIXIE lines similarly extend all the way to the right.}
\label{fig:mu-PBH}
\end{figure}

The fraction of patches (i.e. the initial abundance) which collapse into black holes is determined by
\begin{align}
    \beta &= \int_{\zeta_{\rm c}}^{\infty} \id \zeta P(\zeta) = \frac{1}{2} \rm{erfc}\left( \frac{\zeta_{\rm c}}{\sqrt{2} \sigma}\right)
\end{align}
where ${\rm erfc}(x)$ is the complement of the error function. When a critical density fluctuation re-enters the horizon, that region collapses to form a primordial black hole with some fraction ($\gamma$, which we take to be unity here) of the horizon mass, $M_{\rm pbh} = \gamma M_{\rm H}(z)$. More precisely, the location of the $\delta$-function peak determines the (monochromatic) mass of the PBH distribution via \citep{Nakama2016}
\begin{align}
    M_{\rm pbh} \simeq 10^{9} \, M_{\odot} \, \gamma \left( \frac{g_*}{10.75}\right)^{-1/6} \left( \frac{k_*}{10^2} \, {\rm Mpc}^{-1}\right)^{-2},
\end{align}
scaled here to directly make the link to super-massive black holes apparent.
The initial and late time abundance ($f_{\rm pbh} = \Omega_{\rm pbh}/\Omega_{\rm dm}$) are related through
\begin{align}
    f_{\rm pbh} \simeq  5 \times 10^8\, \gamma^{1/2} \beta \left( \frac{g}{10.75}\right)^{-1/4} \left( \frac{\Omega_{\rm dm}}{0.27}\right) \left( \frac{M_{\rm pbh}}{M_{\odot}} \right)^{-1/2}.
\end{align}
Recalling that for $\delta$ features the induced distortion is $\mu \simeq A_{\zeta}^{\delta} \, W_{\rm \mu} (k_*)$, severe constraints can be mapped onto the PBH parameter space from nondetection by \COBEF. Previous work \citep{Nakama2017} utilized less precise versions of the window functions, so we present updated constraints using the PCA method described above in Fig.~\ref{fig:mu-PBH}. Most notable, the inclusion of information from the $y$ and residual eras can extend previously derived constraints by almost two orders of magnitude.

The results in Fig.~\ref{fig:mu-PBH} show a strong overlap between $y$- and $\mu$-excluded regions. This is primarily due to the fact that a large feature in the PPS is needed to create any primordial black holes (notice how contours extend down to $f_{\rm pbh}\rightarrow 0$). This means for PBH constraints that even the tails of the $y$ and $\mu$ window functions have large constraining power. These tails overlap strongly, unlike the maximal plateaus. The importance of the window function tails also explains why possible gains with \PIXIE, while significant in overall distortion sensitivity, are surprisingly modest for $f_{\rm pbh}$. 

Since both $y$ and $\mu$ can independently constrain wide PBH mass ranges, a simultaneous detection of both distortion parameters (e.g. from \PIXIE) would combine to narrow down a precise mass range for PBHs. A larger amplitude in the $y$ distortion compared to the $\mu$ would signify higher mass PBHs and vice versa.


In Fig.~\ref{fig:mu-PBH} we exclude scales above $k_{\rm rec}\approx 3\times 10^{-2}\;{\rm Mpc}^{-1}$, since these modes enter the horizon after recombination, and the production of SDs is more complex than the window function description given above. It is also noteworthy that beyond scales of $k\sim 1\;{\rm Mpc}^{-1}$ there would also be direct limits on the PPS arising from \Planck.

Finally one important extension not considered here is the formation of PBHs from non-gaussianities, thus moving beyond the simple PPS picture. This is considered in more detail in \cite{Nakama2017} and \citet{Unal2020}, where typically speaking the constraints are weaker for non-gaussian fluctuations due to a lowering of the collapse probability at a given PPS amplitude. In the results this impacts not just the mass ranges probed, but also limits the $f_{\rm pbh}$ visibility in some cases (not all contours stretch to arbitrarily low $f_{\rm pbh}$), which would give \PIXIE a significant boost in constraining power when compared against \COBEF (unlike in Fig.~\ref{fig:mu-PBH}).
In addition, one could expect significant distortion anisotropies to form from the stochastic nature of the dissipation process. These scenarios may be probed with $\mu-T/E$ correlations (see \citet{Rotti2022} for limits from \Planck and \citet{Bianchini2022} for discussions on \COBEF), which can now in principle be computed using a full spectro-spatial descriptions of the CMB anisotropies \citep{Kite2022III}.

\section{Discussion and Conclusions} \label{sec:V}
In this work, we have further elucidated the link between CMB spectral distortions, and exotic models which can source a significant stochastic gravitational wave background. 

After a brief discussion on the generation of spectral distortions from scalar perturbations, we reviewed a simple window function approach to determine the relevant distortion parameters in various approximate schemes (Methods A/B/C/D/PCA). The different window functions can be seen in Fig.~\ref{fig:Window-comp}, and spectral distortion constraints (assuming a $\delta$-function feature) for methods B and PCA are shown in Fig.~\ref{fig:SD-constraints-PPS}. Method B is most widely used in the literature, but misses out on constraining power from some dissipation occurring during the residual era ($5\times10^4 \lesssim z \lesssim 3 \times 10^5$). As a result we recommend using the PCA window function in future works to generate more robust and stringent constraints. 

Spectral distortion limits on scalar dissipation are an integral constraint, which means that they transform (and in general, become stronger) when the shape of a primordial feature transforms away from a $\delta$ function. To appreciate how the constraints evolve, Figs.~\ref{fig:SD-constraints-Gaussian} and \ref{fig:SD-constraints-Box} show limits for Gaussian and Box-shaped features, which have been of recent interest\footnote{Additional cases can be found in \citep{Chluba2012inflaton}}. Perhaps most insightful is the widening of constraints for broad Gaussians, which could dramatically strengthen the GW exclusion plots for SIGWs.

Following this, the formalism from \citet{Kite2020} was presented to compute the $\mu$ distortion from direct dissipation of primordial tensor modes. In analogy to the scalar case, a window function approach was derived and generalized to include GW backgrounds sourced at $z\lesssim 10^8$ when their observational signatures could be altered. A summary of the tensor window functions can be found in Fig.~\ref{fig:tensor-Wmu}, and the corresponding limits seen in Fig.~\ref{fig:SD-constraints-GWs-expanded} highlight the complementary nature of SD constraints in bridging scales between experiments.

From here we moved on to computing the distortion signatures of a non-exhaustive list of exotic models also capable of producing a sizeable SGWB. We first looked at the case of scalar-induced gravitational waves from $\delta$-like features in the primordial power spectrum. Contrary to what is conventionally done in the literature, we performed a mapping which took the constraints on the PPS and translated them into conservative limits on the GW parameter space for these $\delta$ features. A discussion on the subtleties and substructure of these limits is presented in and around Fig.~\ref{fig:SIGW-FIRAS-detail}, while the results of this mapping and the specific impact of spectral distortion constraints are highlighted in Figs.~\ref{fig:SD-SIGW-comparison} and \ref{fig:SIGW-FIRAS-detail-2}. 

A more complete look at the GW landscape for SIGWs (from $\delta$ features) was presented in Fig.~\ref{fig:SD-constraints-GWs-expanded}. For $f \gtrsim 10^{-10}$ Hz, the excluded region is driven by a forest of resonant peaks and has a $1\rightarrow 1$ mapping to PBH constraints on the PPS. Interestingly, in the event that \LiteBIRD makes a detection of primordial tensor modes, a next generation spectral distortion experiment would be capable of determining if SIGWs were the source. The overall shape of these limits will change significantly when considering scalar features other than a $\delta$-function. In general, we expect that for wider features the GW constraints will get stronger, as these shapes will also necessarily widen and deepen the CMB distortion constraints on the PPS. We plan to investigate this in future work.

We additionally rederived the constraints on primordial black holes coming from large enhancements to the power spectrum (assuming Gaussian fluctuations). As an improvement to the previous literature, we utilized the full PCA window function for both the $\mu$ and $y$ distortions, slightly strengthening the constraints (see Fig.~\ref{fig:mu-PBH}). The production of a distribution of PBHs can lead to a significant GW background, particularly in models with large amounts of clustering. CMB spectral distortion signatures of such models are another possible avenue of future research, including the study of their SD anisotropies \citep[e.g.,][]{Kite2022III}.

The opening of the gravitational wave window is a major step forward in our understanding of cosmology, but comes with a number of new challenges that must be addressed. In particular, the ability to confidently disentangle astrophysical and primordial signals will be crucial for exploration of the parameter space of exotic models. CMB spectral distortions provide a powerful multi-messenger probe, complementary to other more direct measurements, that is capable of discriminating such signals.

\section*{Acknowledgements}
We would like to thank Emmanuel Fonseca for discussions relating to the NANOGrav dataset, and Guillem Domènech for helpful insights on SIGWs. 
This project has received funding from the European Research Council (ERC) under the European Union’s Horizon 2020 research and innovation program (grant agreement No 725456; CMBSPEC).
BC was furthermore supported by an NSERC postdoctoral fellowship.
JC was also supported by the Royal Society as a Royal Society University Research Fellow at the University of Manchester.
JCH acknowledges support from NSF grant AST-2108536, NASA grants 21-ATP21-0129 and 22-ADAP22-0145, the Sloan Foundation, and the Simons Foundation.

\section*{Data Availability}

The data underlying in this article are available in this article and can further be made available on request.


\bibliographystyle{mnras}
\bibliography{Lit}



\bsp	
\label{lastpage}
\end{document}